\documentclass{article}

\usepackage[preprint]{neurips_2025}

\usepackage[utf8]{inputenc} 
\usepackage[T1]{fontenc}    
\usepackage{hyperref}       
\usepackage{url}            
\usepackage{booktabs}       
\usepackage{amsfonts}       
\usepackage{nicefrac}       
\usepackage{microtype}      
\usepackage{xcolor}         
\usepackage{amsmath}
\usepackage{amssymb}
\usepackage{mathtools}
\usepackage{amsthm}
\usepackage[linesnumbered,ruled,vlined,algo2e]{algorithm2e}
\usepackage{algorithm}
\usepackage{subfigure}
\usepackage{wrapfig}

\newcommand{\overbar}[1]{\mkern 1.5mu\overline{\mkern-1.5mu#1\mkern-1.5mu}\mkern 1.5mu}

\newcommand{\Lap}{\mathop{\mathrm{Lap}}}
\newcommand{\Vol}{\mathop{\mathrm{Vol}}}

\newcommand{\prob}[1]{\mathbb{P}\left[#1\right]}
\theoremstyle{plain}
\newtheorem{theorem}{Theorem}[section]
\newtheorem{proposition}[theorem]{Proposition}
\newtheorem{lemma}[theorem]{Lemma}

\theoremstyle{definition}
\newtheorem{definition}[theorem]{Definition}
\newtheorem{claim}[theorem]{Claim}

\theoremstyle{remark}

\title{Locally Differentially Private Graph Clustering via \\ the Power Iteration Method}

\author{%
  Vorapong Suppakitpaisarn and Sayan Mukherjee \thanks{Both of the authors contributed to this paper equally.} \\
   The University of Tokyo, Japan\\
  \texttt{vorapong@is.s.u-tokyo.ac.jp, sayan@g.ecc.u-tokyo.ac.jp} \\
}

\begin{document}

\maketitle

\begin{abstract}
  We propose a locally differentially private graph clustering algorithm. Previous works have explored this problem, including approaches that apply spectral clustering to graphs generated via the randomized response algorithm. However, these methods only achieve accurate results when the privacy budget is in $\Omega(\log n)$, which is unsuitable for many practical applications. In response, we present an interactive algorithm based on the power iteration method. Given that the noise introduced by the largest eigenvector constant can be significant, we incorporate a technique to eliminate this constant. As a result, our algorithm attains local differential privacy with a constant privacy budget when the graph is well-clustered and has a minimum degree of \(\tilde{\Omega}(\sqrt{n})\). In contrast, while randomized response has been shown to produce accurate results under the same minimum degree condition, it is limited to graphs generated from the stochastic block model. We perform experiments to demonstrate that our method outperforms spectral clustering applied to randomized response results. 
\end{abstract}

\section{Introduction}
\label{sec:intro}
As the adoption of artificial intelligence expands, ensuring the protection of user privacy has become a critical priority. Various techniques have been proposed to tackle privacy concerns, with differential privacy emerging as a leading approach. Differential privacy, introduced in \citep{dwork2008differential}, quantifies the privacy leakage of a system using a parameter known as the privacy budget. The core idea involves introducing noise to users' data to obscure individual information while still enabling meaningful statistical analysis. The challenge of designing algorithms that can draw accurate insights from this noisy data has garnered significant attention from researchers (surveyed in \citep{zhu2017differentially}), as it is essential to balance privacy protection with the utility of the resulting analysis.

In this work, we focus on a specific variant of differential privacy known as local differential privacy (LDP) introduced in \citep{kasiviswanathan_what_2011}. Unlike traditional differential privacy, which allows data collection before noise is added, LDP requires users to anonymize their data directly on their local devices before transmitting it to a central server. This approach ensures that sensitive information remains protected during transmission, as the data is already corrupted at the source. LDP has been adopted by several major companies \citep{erlingsson_rappor_2014,apple} in their services to safeguard user privacy while still enabling data analysis at scale.

We focus on developing LDP algorithms for social networks, where users are represented as nodes and their relationships as edges. Since these connections are considered sensitive, they are protected using privacy notions such as edge LDP introduced in \citep{qin2017generating} or node LDP introduced in \citep{ye2020towards}. However, with some exceptions like \citep{zhang2020differentially}, node LDP is generally too stringent, making it difficult to release useful information in most applications. As a result, as discussed in \citep{imola_locally_2021}, the majority of research in LDP has centered around the more practical edge LDP framework.

To protect user's information, one widely used technique is randomized response, also known as edge flipping discussed in \citep{warner_randomized_1965, mangat_improved_1994, wang2016using}. In this method, before a user sends a bit vector which encodes their list of friends to a central server, each bit in the vector is flipped with a certain probability. The server aggregates the obfuscated adjacency vector to construct an obfuscated version of the graph. Although it is possible to compute various graph statistics from this obfuscated data, the accuracy of these statistics is often reduced. Algorithms designed specifically to publish particular statistics such as \citep{imola_locally_2021,imola_communication-efficient_2022} tend to offer more precise and insightful results about the graph.

Graph clustering illustrates how analyzing a graph obfuscated by randomized response can lead to inaccurate results. Let $n$ be the number of nodes in the input graph. In \citep{hehir2022consistent}, the authors demonstrated that spectral clustering discussed in \citep{ng2001spectral} can yield accurate results with a privacy budget in $O(1)$, provided the input graphs are generated from stochastic block models (introduced in \citep{holland1983stochastic} and have an average and minimum degree of $\Theta(\sqrt{n})$. For general graphs, \citep{mukherjee2023robustness} showed that applying spectral clustering to randomized response data only yields accurate results when the privacy budget $\epsilon \in \Omega(\log n)$, which is too large for many real-world applications. 

Although numerous algorithms have been proposed for clustering under differential privacy (e.g. \citep{ji2020community, mohamed2022differentially, chen2023private, imola2023differentially, epasto2024power,hedifferentially}), relatively few have been developed specifically for publishing clustering results under edge LDP. Aside from the work mentioned in the previous paragraph, the only other algorithm we are aware of is \citep{fu2023gc}. The work targets node LDP rather than edge LDP.

\subsection{Our contributions}

In this work, we aim to develop a dedicated algorithm for graph clustering under the edge LDP framework. Rather than using non-interactive methods like the edge flipping algorithm, we propose an interactive approach, which has been shown in \citep{henzinger2024tighter, hillebrand2024cycle} to achieve better performance for many edge LDP tasks.

Specifically, we draw inspiration from the work in \citep{betzerpublishing}, where the authors employ multi-round interactive algorithms to compute iterative matrix multiplications for Katz centrality. Spectral clustering can also be derived through iterative matrix multiplication using the Power Iteration Clustering (PIC) algorithm proposed in \citep{PIC-LinCohen-2010,PIC-Provably-BoustidisKambadurGittens-2015,wang2020nearly}. We hence propose extending this approach to calculate clusters via the PIC algorithm under the edge LDP framework. An independent work \cite{koskela2025pricedifferentialprivacyspectral} proposes an iterative clustering algorithm under differential privacy; however, it does not address the LDP setting, and its analysis is limited to graphs generated from the SBM.

Unfortunately, calculating the PIC algorithm under the edge LDP framework is not straightforward. While the goal is to compute the second eigenvector through the iterative process, the largest component of the result is the first eigenvector. In a non-private setting, the first eigenvector, being a uniform vector, does not interfere with the calculation of the PIC algorithm. However, when protecting users' sensitive information under edge LDP, noise must be added at a magnitude comparable to the largest terms. This causes the noise to dominate the result, especially as the number of iterations increases, leading to a significant loss in accuracy.

We propose a technique to eliminate the largest constant term, enabling the development of an algorithm that achieves accurate results with a constant privacy budget when the minimum degree of the input graph is \(\tilde{\Omega}(\sqrt{n})\). Recall that randomized response is proven to yield accurate results for graphs generated by the stochastic block model when the minimum degree is \(\tilde{\Omega}(\sqrt{n})\). Our algorithm, however, provides precise results under the same minimum degree condition but applies to general graphs, not limited to those generated by the model. This extends the applicability of our clustering algorithm to a wider range of input graphs.


Our algorithm is computationally efficient. It requires $O(\log n)$ interactions between users and the central server, with each node having a computational complexity of $O(n)$ per iteration. The central server also has a computational complexity of $O(n)$ per iteration. Consequently, the total computation time of our algorithm is $O(n \log n)$. Additionally, the communication cost for each user is also $O(n \log n)$.

Compared to the spectral clustering algorithm applied to the randomized response results in \citep{hehir2022consistent, mukherjee2023robustness}, our iterative method is significantly more memory-efficient. In the previous approach \citep{imola_communication-efficient_2022,hillebrand2023unbiased}, the server requires $\Theta(n^2)$ bits of memory to store the randomized response results. In contrast, our algorithm reduces the memory requirement to $\Theta(n)$ for both the server and the users. This improvement enables our method to handle graphs with a large number of nodes, which would be infeasible to process using the earlier algorithm.

We validate our algorithm through experiments on graphs generated using the stochastic block model \citep{holland1983stochastic} and the Reddit graph \citep{hamilton2017inductive}. Compared to applying the spectral clustering algorithm to the randomized response results \citep{hehir2022consistent}, our algorithm produces clustering results that are closer to those of the original spectral clustering algorithm in almost all cases. Notably, there are several instances where the previous algorithm yields random outcomes, while our algorithm consistently produces results identical to the original spectral clustering.

\section{Preliminaries}
\label{sec:prelim}

\subsection{Notation}
\label{subsec:notation}

Throughout this paper, we consider a graph \( G = (V, E) \) with \( n \) vertices. 
Let \( S \subseteq V \) represent a subset of vertices, and \( \overbar{S} \) denote its complement \( V \setminus S \). 

Let \( S \) and \( S' \) be two disjoint subsets of \( V \) (meaning \( S \cap S' = \varnothing \)). We denote by \( e_G(S, S') \) the number of edges in \( G \) that have one endpoint in \( S \) and the other in \( S' \). For each subset \( S \subseteq V \), let \( \mathrm{Vol}_G(S) \) denote the number of edges with both endpoints in \( S \). We refer to \( \mathrm{Vol}_G(S) \) as the \textit{volume} of \( S \).
For \( S, S' \subseteq V \), the quantity
$d_{\text{vol}}(S, S')$ is defined as $\min ( \mathrm{Vol}_G(S \triangle S') + \mathrm{Vol}_G(\overbar{S} \triangle \overbar{S'}),\mathrm{Vol}_G(S \triangle \overbar{S'}) + \mathrm{Vol}_G(\overbar{S} \triangle S') ).$
Since \( S \triangle S' = \overbar{S} \triangle \overbar{S'} \), this simplifies to \( d_{\text{vol}}(S, S') = \min \left( 2 \mathrm{Vol}_G(S \triangle S'), 2 \mathrm{Vol}_G(S \triangle \overbar{S'}) \right) \). Two cuts \( (S, \overbar{S}) \) and \( (S', \overbar{S'}) \) are considered similar if \( d_{\text{vol}}(S, S') \) is small. We also define the \textit{normalized discrepancy} as
    $d_{\rm norm}(S,S') = \frac{d_{\rm vol}(S,S')}{{\rm Vol}_G(V)}$. 
Given that \( d_{\rm vol}(S, S') \leq {\rm Vol}_G(V) \), normalization ensures that \( 0 \leq d_{\rm norm}(S, S') \leq 1 \). When \( S \) is fixed and nodes are randomly assigned to \( S' \) with uniform probability, \( d_{\rm norm}(S, S') \) tends to be close to 1.

Any real symmetric \( n \times n \) matrix \( A \) has \( n \) real eigenvalues. We denote the \( i \)-th smallest eigenvalue of \( A \) as \( \lambda_i(A) \), so that \( \lambda_1(A) \geq \lambda_2(A) \geq \cdots \geq \lambda_n(A) \). The eigenvector corresponding to $\lambda_i(A)$ is denoted by $\mathbf{v}_i(A) = [\nu_{i,1}, \dots, \nu_{i,n}]^\intercal$.

For each \( i \in [1, n] \), let \( a_i = [a_{i,1}, \ldots, a_{i,n}]^\intercal \) represent the adjacency list of user \( v_i \), where \( a_{i,j} = 1 \) signifies the existence of an edge between \( v_i \) and \( v_j \) (i.e., \( (v_i, v_j) \in E \)), and \( a_{i,j} = 0 \) indicates no edge. The degree of node \( v_i \), denoted by \( d_i \), reflects the number of edges connected to \( v_i \). In the context of a locally differentially private algorithm, it is assumed that each user \( v_i \) is aware only of their own adjacency vector \( a_i \), which contains sensitive personal information.

\subsection{Edge local differential privacy}

We define two adjacency lists, \(a\) and \(a'\), as neighboring if they differ by exactly one bit, meaning that one can be transformed into the other by either adding or removing a single edge connected to node \(v_i\). The concept of edge local differential privacy is formalized as follows:

\begin{definition}[$\epsilon$-Edge LDP Query]
\label{ref1}
Let \(\epsilon > 0\). A randomized query \(\mathcal{R}\) is said to satisfy \(\epsilon\)-edge local differential privacy ($\epsilon$-edge LDP) if, for any pair of neighboring adjacency lists \(a\) and \(a'\), and any possible outcome set \(S\), 
    $\prob{\mathcal{R}(a) \in S} \leq e^{\epsilon} \prob{\mathcal{R}(a') \in S}$.
\end{definition}

\begin{definition}[$\epsilon$-edge LDP Algorithm \citep{qin2017generating}]
\label{def:ldp-algorithm}
An algorithm \(\mathcal{A}\) is said to be \(\epsilon\)-edge LDP if, for any user \(v_i\), and any sequence of queries \(\mathcal{R}_1, \dots, \mathcal{R}_\kappa\) posed to user \(v_i\), where each query \(\mathcal{R}_j\) satisfies \(\epsilon_j\)-edge local differential privacy (for \(1 \leq j \leq \kappa\)), the total privacy loss is bounded by \(\epsilon_1 + \dots + \epsilon_\kappa \leq \epsilon\).
\end{definition}
If an algorithm \(\mathcal{A}\) is \(\epsilon\)-edge LDP, it is also said to have a privacy budget of \(\epsilon\).
Next, we introduce a query that satisfies \(\epsilon\)-edge LDP which designed to estimate a real-valued statistic based on the adjacency vector.
\begin{definition}[Edge Local Laplacian Query \citep{hillebrand2023unbiased}] \label{def:lap}
    Let \( f: \{0,1\}^n \rightarrow \mathbb{R} \) be a function defined on adjacency lists, and let \( a \sim a' \) represent neighboring adjacency lists. The global sensitivity of \( f \), denoted as \( \Delta_f \), is defined as:
    $\Delta_f = \max_{a \sim a'} |f(a) - f(a')|$.
    For any \(\epsilon > 0\), a query that returns \( f(a) + \Lap(\Delta_f / \epsilon) \) is \(\epsilon\)-edge LDP. Here, \(\Lap(b)\) refers to noise sampled from the Laplace distribution with scale parameter \( b \).
    \label{laplacian}
\end{definition}

\subsection{Spectral clustering}

For a given graph \( G \), the primary objective of clustering techniques is to identify a cut \((S, \overbar{S})\) such that the number of edges crossing between \( S \) and \( \overbar{S} \), denoted by \( e_G(S, \overbar{S}) \), is minimized, while most of the edges are concentrated within \( S \) or \( \overbar{S} \). To avoid trivial cuts (such as when \( S \) contains only a single vertex), it is common to define the \emph{conductance}, \( \phi_G(S) = e_G(S, \overbar{S}) / \min\{\mathrm{Vol}_G(S), \mathrm{Vol}_G(\overbar{S})\} \), and seek cuts that minimize \( \phi_G(S) \) \citep{shi2000normalized}. The conductance of the graph, denoted by \( \phi(G) \), is given by
$\phi(G) = \min_{\varnothing \subsetneq S \subsetneq V} \phi_G(S)$.
Unless otherwise stated, we use \( S^\ast \) to denote the subset that achieves the minimum normalized cut, where \( \phi_G(S^\ast) = \phi(G) \).

Let $B = (b_{i,j})_{1 \leq i,j \leq n}$ be the transition-probability matrix of a random walk on $G$, given by $b_{i,i} = 0$ for all $i$ and $b_{i,j} = a_{i,j}/d_i$ for all $i \neq j$. We have that $-1 \leq \lambda_i(B) \leq 1$ for all $i$, $\lambda_1(B) = 1$, and $\mathbf{v}_1(B) = {\textstyle[\frac1{\sqrt{n}}, \frac1{\sqrt{n}}, \dots, \frac1{\sqrt{n}}]^\intercal}$.

Observe that when \( I \) is the identity matrix, the matrix \( I - B \) is referred to as the \textit{random walk normalized Laplacian matrix} \citep{von2007tutorial}. The eigenvectors of \( I - B \) are identical to those of \( B \). More specifically, it is known that, for all $i$, \( \mathbf{v}_i(I - B) = \mathbf{v}_{n-i}(B) \).

The spectral clustering algorithm \citep{shi2000normalized} computes the eigenvector \(\mathbf{v}_2(B) = [\nu_1, \dots, \nu_n]^\intercal\), and then produces the cut \(S' = \{v_i : \nu_i > 0\}\) as the clustering result. Since \(\phi_G(S') \leq 2 \sqrt{\phi_G(S^*)}\) \citep{alon1986eigenvalues}, it is established that the cut produced by the spectral clustering algorithm achieves a low conductance. Additionally, according to \citep{peng2015partitioning}, we have \(d_{\mathrm{vol}}(S', S^*) = O\left(\frac{\phi(G)}{\lambda_3(B)} \cdot \mathrm{Vol}_G(V)\right)\), indicating that \(S'\) closely approximates \(S^*\) in a graph that is well-clustered.

The normalized Laplacian matrix \( L = (\ell_{i,j})_{1 \leq i,j \leq n} \), defined by \( \ell_{i,j} = -a_{i,j}/\sqrt{d_i \cdot d_j} \) for \( i \neq j \) and \( \ell_{i,i} = 1 \), is commonly used in spectral clustering algorithms that aim to minimize the conductance. However, in this work, we opt for the random walk normalized Laplacian matrix, as calculating spectral clustering under the normalized Laplacian is more complex in the edge  LDP setting. Notably, when the desired number of clusters is two, the results of spectral clustering using the random walk normalized Laplacian matrix are at least as good as those obtained with the normalized Laplacian matrix \citep{von2007tutorial}.

\subsection{Power iteration clustering}
\label{subsec:PIC}

While spectral clustering can produce a cut with a small cut-ratio, it requires computing the eigenvector \(\mathbf{v}_2(B)\), which can be computationally expensive. To address this, the power iteration clustering algorithm \citep{PIC-LinCohen-2010} offers a more efficient method for estimating the eigenvector, significantly reducing the computation time.

Let \(\mathbf{x}\) be a vector of length \(n\) where each element is independently drawn from a Gaussian distribution. It is known that \(\mathbf{x}\) can be expressed as \( c_1 \lambda_1(B) \mathbf{v}_1(B) + \dots + c_n \lambda_n(B) \mathbf{v}_n(B) \), where \(c_1, \dots, c_n\) are independent random variables also drawn from a Gaussian distribution. Therefore, for a sufficiently large $T$, applying \(B^T\) to \(\mathbf{x}\) gives 
\begin{eqnarray} B^T \mathbf{x} & = & c_1 \lambda_1(B)^T \mathbf{v}_1(B) + \dots + c_n \lambda_n(B)^T \mathbf{v}_n(B) \nonumber \\ & = & c_1 \cdot {\textstyle\left[\frac1{\sqrt{n}}, \frac1{\sqrt{n}}, \dots, \frac1{\sqrt{n}}\right]^\intercal} + c_2 \lambda_2(B)^T \mathbf{v}_2(B) + \dots +c_n \lambda_n(B)^T \mathbf{v}_n(B).
\label{eqn1}
\end{eqnarray} 
When \(\lambda_3(B) \ll \lambda_2(B)\), the term \(B^T \mathbf{x}\) is approximately 
$B^T \mathbf{x} \approx c_1 {\textstyle\left[\frac1{\sqrt{n}}, \frac1{\sqrt{n}}, \dots, \frac1{\sqrt{n}}\right]^\intercal} + c_2 \lambda_2(B)^T \mathbf{v}_2(B),$
meaning the order of elements in \(B^T \mathbf{x}\) closely mirrors that of \(\mathbf{v}_2(B)\). Therefore, clustering can be performed using \(B^T \mathbf{x}\) instead of \(\mathbf{v}_2(B)\), yielding results similar to those from the spectral clustering algorithm.


\subsection{Assumptions} \label{subsec:assumption}

(1) The minimum degree is at least $2\sqrt{n} \log^4 n$:  The first assumption is essential for any graph clustering algorithm under edge LDP with a constant privacy budget. Protecting the connections of low-degree nodes requires adding so much noise that their contributions are obscured, resulting in unstable clustering outcomes for these nodes.

(2) There exists a constant $g$ such that for all $i \geq 3$, $\lambda_i(B) + 1 \leq \frac{\lambda_2(B) + 1}{g}$:  The second assumption is a standard prerequisite for iterative spectral clustering algorithms. This assumption ensures the convergence of the iterative process. A comprehensive explanation supporting this assumption is provided in \citep{PIC-Provably-BoustidisKambadurGittens-2015}.

(3) There exists $\delta \approx 1$ and $\gamma < 1$ such that the components of $\mathbf v_2(B)$ satisfies $\left|\left\{i : |\nu_i| \geq \frac{\gamma}{\sqrt{n}}\right\}\right| \geq \delta \cdot n$: We demonstrate in Appendix \ref{appendix:eigenvector} that the third assumption holds when the graph is well-clustered and most nodes have a degree cluster close to the average degree of the cluster to which they belong. We observe that the graphs generated by the stochastic block model have this property.
In addition to our mathematical proof in the appendix, it is empirically demonstrated in \citep{abbe2020entrywise} that most of the values in the eigenvectors is in $\Theta(1/\sqrt{n})$.  Additionally, \citep{balakrishnan2011noise} shows that this assumption can be satisfied when \( B \) is a node similarity matrix with certain additional properties. 

(4) The number of nodes $n$ is larger than a constant $C$:  The final assumption is a common requirement for most differentially private algorithms. A large user base typically allows the added noise, introduced to protect sensitive information, to average out in the results.

We discuss in Appendix \ref{appendix:assumption} that our assumption is at least as strong as those in the previous works \cite{hehir2022consistent}.

\section{Our algorithm} \label{sec:alg}

\begin{algorithm}[t]
    \caption{Private power iteration clustering \label{alg:private_modified_PIC}}
    \KwIn{Graph $G = (V,E)$ where $V = \{v_1, \dots, v_n\}$ and its adjacency matrix is $A = (a_{i,j})_{1 \leq i,j \leq n}$, privacy budget $\epsilon$, number of iterations $T = \frac{2 \log n}{\log g}$, clipping factor $\mathsf{c}$, and parameter $\zeta = \frac{1}{n}$}
    \KwOut{A cut of $G$ denoted by $S \subset V$}
    \textbf{[User $i$]} Compute the degree of \( v_i \), denoted by \( d_i \). Broadcast \( \tilde{d}_i \gets d_i + \Lap(10/\epsilon) \) to all users and the server. \;
    
    \textbf{[Server]} Calculate $\delta \gets \min_i \tilde{d_i} - \frac{10}{\epsilon} \log \frac{n}{2\zeta}$. Broadcast $\delta$ to all users.

    \textbf{[User $i$]} If \(d_i < \delta\), randomly select \(j\) such that \(a_{i,j} = 0\), then set \(a_{i,j} = 1\) and increment \(d_i\) by one. Repeat this process until \(d_i \geq \delta\).
    
    \textbf{[Server]} Initiate the vector $\mathbf{x}^{(0)} = [x^{(0)}_1, \dots, x^{(0)}_n]^\intercal$ where $x^{(0)}_i$ is chosen from the Gaussian distribution with expected value $0$ and standard deviation $1$. Broadcast the vector $\mathbf{x}^{(0)}$ to all users.\;
    
    \For{$t=1,\ldots, T$}{
    \textbf{[User $i$]} Calculate $w_i^{(t)} = \frac{1}{2} x_i^{(t - 1)} + \frac{1}{2} \sum_j a_{i,j} \frac{x_j^{(t - 1)}}{d_i} - \frac{1}{n} \sum_j x_j^{(t - 1)} + \Lap\left(\frac{5 \cdot T}{9 \cdot \epsilon} \max_j \frac{|x_j^{(t - 1)}|}{\delta}\right).$ 
    
    \textbf{[User $i$]} Let $U = \mathsf{c}\cdot \frac{5 \cdot T}{9 \cdot \epsilon} \max_j \frac{|x_j^{(t - 1)}|}{\delta}$, also let $x_i^{(t)} = U$ if $w_i^{(t)} > U$, $x_i^{(t)} = -U$ if $w_i^{(t)} < -U$, and $x_i^{(t)} = w_i^{(t)}$ otherwise. Calculate and send $x_i^{(t)}$ to the server.

    \textbf{[Server]} Aggregate the values \(x_i^{(t)}\) into a vector \(\mathbf{x}^{(t)} = [x_1^{(t)}, \dots, x_n^{(t)}]^\intercal\), and broadcast this information to all users.
    }
    \textbf{[Server]} Return $S \gets \{v_i : \mathbf{x}^{(T)}_i > 0\}$.
\end{algorithm}

We present our algorithm in Algorithm~\ref{alg:private_modified_PIC}. As shown in Lines 6--7, the computation closely resembles the iterative update $\mathbf{x}^{(t)} = B \cdot \mathbf{x}^{(t-1)}$, ultimately leading to $\mathbf{x}^{(T)} = B^T \cdot \mathbf{x}^{(0)}$. However, our update includes four modifications at Line 6: (1) the replacement of the iterative update $\mathbf{x}^{(t)} = B \cdot \mathbf{x}^{(t-1)}$ with the lazy random walk $\mathbf{x}^{(t)} = \frac{1}{2}I + \frac{1}{2} B \cdot \mathbf{x}^{(t-1)}$; (2) the addition of the term $-\frac{1}{n} \sum_j x_j^{(t-1)}$; (3) the addition of the Laplace noise $\Lap\left(\frac{5T}{9\epsilon} \max_j \frac{|x_j^{(t-1)}|}{\delta} \right)$ to ensure privacy; and (4) at Line~7, we apply clipping such that $|w_i^{(t)}| \leq U$. The Laplacian noise is introduced to preserve users' privacy, while the clipping operation helps control the influence of extreme values. On the other hands, we want to highlight the first and second change in the subsequent paragraphs.

\textbf{Key contribution 1 : Elimination of the leading eigenvector} The inclusion of the additional term $-\frac{1}{n} \sum_j x_j^{(t - 1)}$ constitutes the key contribution of this work. Recall Equation~(\ref{eqn1}). Since \(\lambda_2(W) < 1\), the term \(\alpha_2 \lambda_2(W)^T \mathbf{v}_2(W)\) diminishes compared to the leading term as \(T\) increases. On the other hand, the size of the Laplace noise added depends on the largest element of \(\mathbf{x}^{(t-1)}\), which is determined by the leading term. Hence, for larger \(T\), the noise magnitude dominates over the term \(\alpha_2 \lambda_2(W)^T \mathbf{v}_2(W)\). This causes \(\mathbf{x}^{(T)}\) to deviate significantly from \(\mathbf{v}_2(W)\), reducing the accuracy of the results.

To address this, we introduce the matrix \(\tilde{W} = (\tilde{w}_{i,j})_{1 \leq i,j \leq n}\), where \(\tilde{w}_{i,j} = w_{i,j} - 1/n\) for all $i,j$. We show in Appendix \ref{appendix:elimination} that for all \(i \geq 1\), \(\lambda_i(\tilde{W}) = \lambda_{i+1}(W)\) and \(\mathbf{v}_n(\tilde{W}) = \mathbf{v}_1(W)\). Additionally, \(\mathbf{v}_n(\tilde{W}) = \mathbf{v}_1(W) = {\textstyle[\frac1{\sqrt{n}}, \frac1{\sqrt{n}}, \dots, \frac1{\sqrt{n}}]^\intercal}\) and \(\lambda_n(\tilde{W}) = 0\).

With this update, the leading term \(\alpha_1 \cdot {\textstyle[\frac1{\sqrt{n}}, \frac1{\sqrt{n}}, \dots, \frac1{\sqrt{n}}]^\intercal}\) from (\ref{eqn1}) is eliminated. The term \(\alpha_2 \lambda_2(\tilde{W})^T \mathbf{v}_2(\tilde{W})\) now becomes the leading term, and we can ensure that the Laplace noise (the fourth term of Line 6 in Algorithm \ref{alg:private_modified_PIC}) is substantially smaller than the new leading term. The subtraction of the third term in the calculation at Line 6 reflects the update from \(W\) to \(\tilde{W}\).

\textbf{Key contribution 2: Replacing the random walk with a lazy random walk} Recall that all eigenvalues of the matrix \(B\) lie between $1$ and $-1$. In certain networks, such as bipartite graphs, \(\lambda_n(B)\) can be close to $-1$. This causes the final term in Equation~(\ref{eqn1}) to oscillate, preventing the calculation of \(B^T\mathbf{x}\) from converging. To address this, we propose replacing \(B\) with \(W = \frac{1}{2} I + \frac{1}{2} B\). Note that for all \(i\), \(\mathbf{v}_i(W) = \mathbf{v}_i(B)\) and \(\lambda_i(W) = \frac{1}{2} \lambda_i(B) + \frac{1}{2}\). Consequently, for all \(i\), \(0 \leq \lambda_i(W) \leq 1\). By the second assumption in Section \ref{subsec:assumption}, which is \(\lambda_i(W) \leq \frac{\lambda_2(W)}{g}\) for all $i \geq 3$, we do the approximation discussed in Section \ref{subsec:PIC} even when some \(\lambda_i(B)\) are negative. This modification leads to the first two terms of the calculation in Line 6.

\subsection{Properties of our algorithm}
\label{sec:proofs}

\textbf{Privacy} We show in Appendix \ref{sec:privacy} that our algorithm is $\epsilon$-edge LDP. While our precision results need the assumptions defined in Section \ref{subsec:assumption}, the privacy result works for any graph and any parameter setting.
 
\textbf{Computation time} The primary computational bottleneck of Algorithm~\ref{alg:private_modified_PIC} occurs in Line 6. In this step, the per-node computational complexity for each iteration is \(O(n)\). To achieve accurate results, the required number of iterations \(T\) is given by \(2\frac{\log n}{\log g} = \Theta(\log n)\), leading to an overall computational complexity of \(O(n \log n)\) per node. In contrast, the central server has minimal computational demands. Its responsibilities are limited to generating the initial vector, receiving calculation results, and distributing them to all users.

\textbf{Communication cost} While each user uploads only one real number \(x_i^{(t)}\) to the server at each iteration, they must download the entire vector \(\mathbf{x}^{(t)}\) in Line 8 of the algorithm. This results in a total communication cost of \(O(n \log n)\) for each user.

\textbf{Memory consumption} During iteration \( t \), the central server and all users only need to store two vectors: \( \mathbf{x}^{(t-1)} \) and \( \mathbf{x}^{(t)} \). As a result, the memory consumption for all parties is \( O(n) \). 
This is a significant improvement compared to the randomized response method. Even for sparse input graphs, the randomized response mechanism flips each relationship with a constant probability, leading to a graph with \( \Omega(n^2) \) edges. Storing such a graph, with \( \Omega(n^2) \) edges, requires a prohibitive amount of memory on the server, making it infeasible to design an LDP algorithm for large input graphs \citep{imola_communication-efficient_2022}. In contrast, our approach requires only \( O(n) \) memory, enabling our algorithms to handle input graphs with millions of nodes efficiently.


\textbf{Precision}
\label{subsec:precision} The precision analysis is the most technical part of this paper. Our analysis include the results in Appendix \ref{appendix:minimumDegree}, which show that when the value $\delta$ is a precise estimation of the minimum degree of $G$. If the minimum degree of $G$ is $\tilde{\Omega}(\sqrt{n})$, then the estimation $\delta$ is also in $\tilde{\Omega}(\sqrt{n})$ with large probability. Then, in Appendix \ref{appendix:sizeofnoise}, we show that, when $\delta$ is $\tilde{\Omega}(\sqrt{n})$, the size of Laplace noise at Line 6 is small. Specifically, we demonstrate that the noise scale, given by $\frac{5T}{9\epsilon} \max_j \frac{|x_j^{(t - 1)}|}{\delta}$, is negligible compared to the magnitude of $\mathbf{x}^{(t)}$. Consequently, the noise term $\mathbf{y}^{(t)}$ does not dominate the calculation. 

Let $\mathbf{v}_j(\tilde{W}) = [v_{j,1}, \dots, v_{j,n}]^\intercal$ be the $j$'th eigenvector of $\tilde{W}$, and let \( c_1, \dots, c_n \in \mathbb{R} \) be coefficients such that \( \mathbf{x}^{(0)} = \sum_{j=1}^n c_j \mathbf{v}_j(\tilde{W}) \). Additionally, for all \( t \), suppose the noise added during iteration \( t \) of the algorithm is \( \mathbf{y}^{(t)} \), and that \( e_1^{(t)}, \dots, e_n^{(t)}\in \mathbb R \) are coefficients such that \( \mathbf{y}^{(t)} = \sum_{j=1}^n e_j^{(t)} \mathbf{v}_j(\tilde{W}) \). In Lemma \ref{lem:e}, we show that \(x_i^{(T)} = \sum_{j = 1}^n \tilde{c_j} v_{j,i},\) where $\tilde{c_j}$ is given by $ \tilde{c_j} = c_j \lambda_j(\tilde{W})^T + \sum_{t = 1}^T e_j^{(t)} \lambda_j(\tilde{W})^{T - t}$.
In Theorem \ref{thm:precision}, we demonstrate that 
$\left|c_1 \lambda_1(\tilde{W})^T v_{1,i}\right| > \left|\sum_{t=1}^T e_1^T \lambda_1(\tilde{W})^{T-t} v_{1,i} + \sum_{j=2}^n \tilde{c}_j v_{j,i}\right|$
with probability at least \( 0.95 - o(1) \). The term \( c_1 \lambda_1(\tilde{W})^T v_{1,i} \) dominates and determines the sign of \( x_i^{(T)} \). 

Since \( \lambda_1(\tilde{W})^T \) is positive, we conclude that when \( c_1 v_{1,i} > 0 \), \( x_i^{(t)} > 0 \) with high probability. Recall that the outcome of the spectral clustering algorithm is \( S_{\rm orig} = \{v_i : v_{1,i} > 0\} \). Thus, when \( c_1 > 0 \), the result \( S_{\rm alg} \) closely resembles \( S_{\rm orig} \) with high probability. Conversely, when \( c_1 < 0 \), the result \( S_{\rm alg} \) is similar to \( V \setminus S_{\rm orig} \) with high probability. Therefore, our algorithm is likely to produce a small \( d_{\rm vol}(S_{\rm alg}, S_{\rm orig}) \). In conclusion, the results are comparable to those obtained from the spectral clustering algorithm.



\section{Experimental results}
\label{sec:exp}

For all experiments, we use the normalized discrepancy \( d_{\rm norm} \), as defined in Section \ref{subsec:notation}, to assess precision. Remember that when the normalized discrepancy is small, the outcome closely resembles that of the original spectral clustering algorithm, indicating a high-quality clustering result. The reported values represent the average of 10 experiments, which we consider sufficient, as the variance in precision across each set of experiments is typically small. We show some of the results on graphs generated from the stochastic block model and real graphs on this section. The remaining results can be found in Appendix \ref{appendix:exp}.

\subsection{Results on graphs generated from the stochastic block model}

Our first set of experiments are conducted on graphs generated using the stochastic block model (SBM) \citep{holland1983stochastic}. This model is chosen because it ensures that the generated graphs are well-clustered and consist of exactly two clusters. Furthermore, SBM has been widely employed in prior studies to analyze spectral clustering under local differential privacy \citep{hehir2022consistent}. In this model, the set of \(n\) nodes is divided into two clusters of sizes \(n_1\) and \(n_2\), where \(n_1 + n_2 = n\). Two nodes within the same cluster are connected with probability \(p\), while nodes from different clusters are connected with probability \(q\). While in most cases \(p \gg q\), this paper also considers the scenario where \(q > p\). Unless otherwise specified, we set \( n = 10,000 \), \( n_1 = n_2 = 5,000 \), \( p = 0.3 \), \( q = 0.2 \), the clipping factor $\mathsf{c} = 10$, and the privacy budget \( \epsilon = 1 \). 

The value of \( n \) is chosen to be \( 10,000 \) due to the memory requirements of the benchmark algorithm, randomized response, which requires \( \Omega(n^2) \) bits to store the entire graph for spectral clustering calculations. 
We believe that graphs of this size are sufficient to effectively demonstrate the empirical properties of our algorithm. Given the constraints of our local computational environment, handling larger graphs is not feasible. 
We select \( p = 0.3 \) and \( q = 0.2 \) because these values are close enough to highlight the precision of our algorithm in distinguishing clusters. 
We set the clipping factor $\mathsf{c} = 10$, as it is the integer closest to $\log n \log g$ for well-clustered graphs generated using the stochastic block model. Recall that, when $\mathsf{c} = \log n \log g$, the clipping is applied only with small probability.
 The privacy budget is set to \( \epsilon = 1 \) as it is a standard value commonly used in experiments of other local differential privacy algorithms \citep{hillebrand2023unbiased}.

To the best of our knowledge, only one graph clustering algorithm under local differential privacy has been explored in the literature. This algorithm employs the spectral clustering method on graph processed using randomized response \citep{hehir2022consistent}. Therefore, we select this algorithm as the benchmark for our study.

\begin{figure}[t]
    \centering
    \subfigure[Comparison across different privacy budget]{%
        \includegraphics[width=0.28\textwidth]{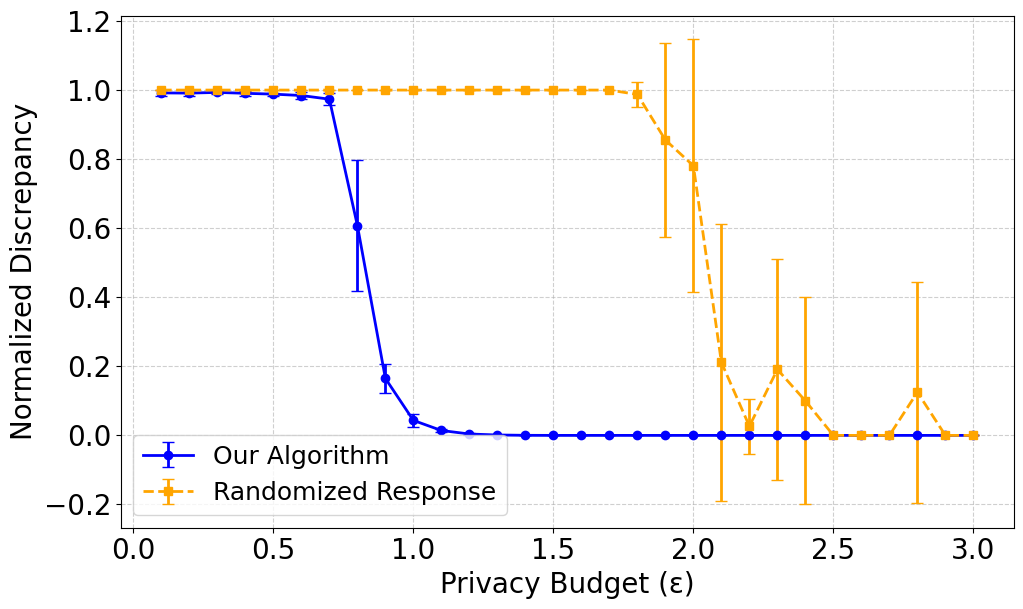} 
        \label{subfig:eps}
    }
    \hfill
    \subfigure[
Comparison across different graph sizes]{%
        \includegraphics[width=0.28\textwidth]{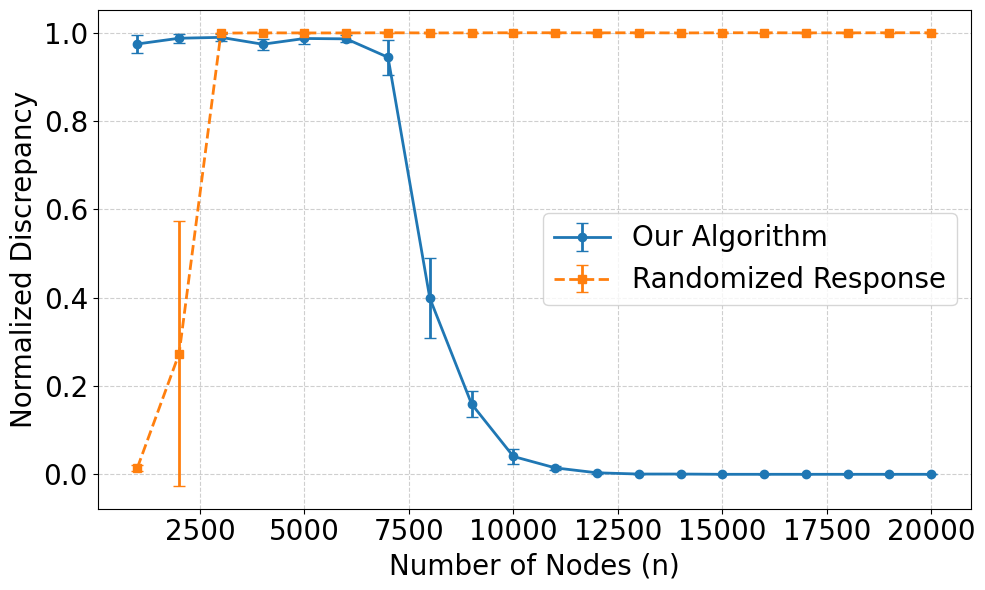} 
        \label{subfig:n}
    }
       \subfigure[Comparison across different graph density when $\epsilon = 1$]{%
        \includegraphics[width=0.19\textwidth]{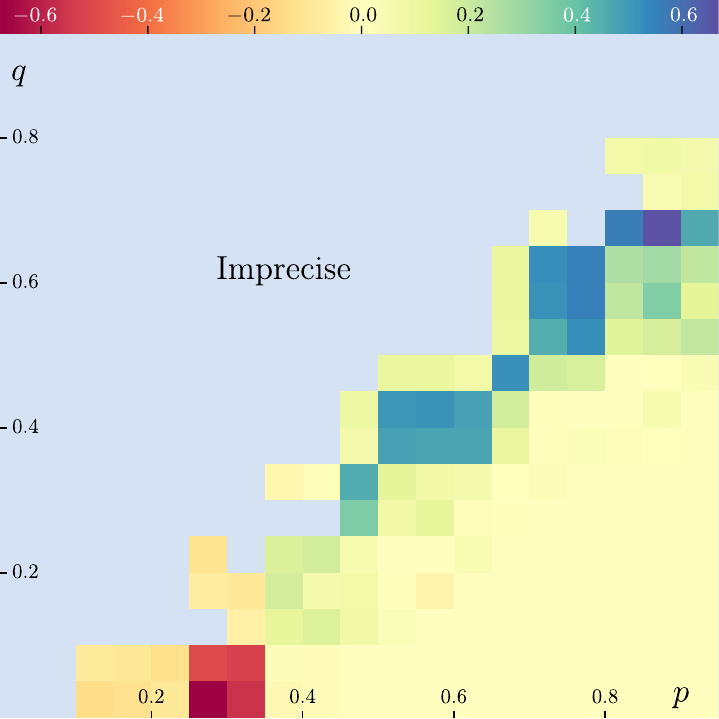} 
        \label{subfig:density1}
    }
    \hfill
    \subfigure[
Comparison across different graph density when $\epsilon = 1.5$]{%
        \includegraphics[width=0.19\textwidth]{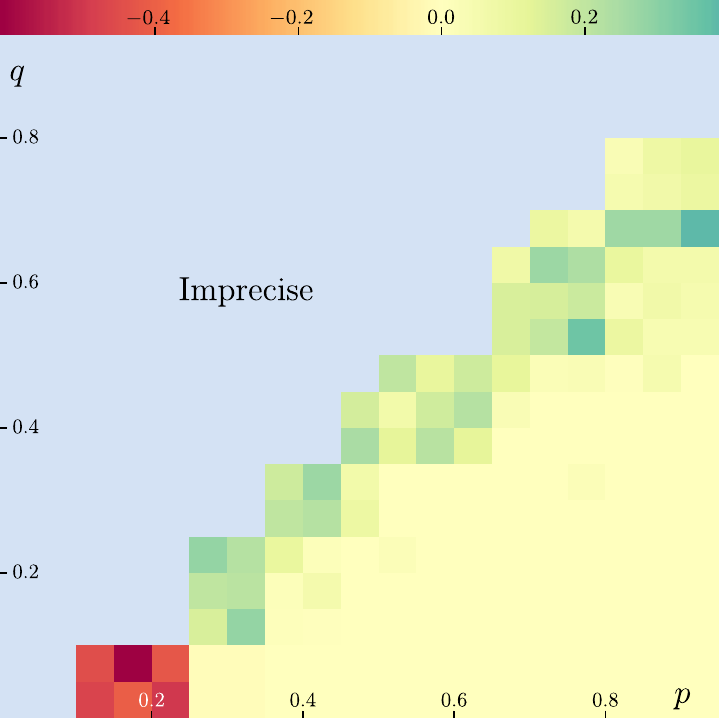} 
        \label{subfig:density15}
    }
    \caption{
Comparison of the normalized discrepancy between our algorithm and the randomized response-based algorithm on the graphs generated from the stochastic block model. The results shown in Figures \ref{subfig:density1} and \ref{subfig:density15} represent the differences in the normalized discrepancies between the two algorithms.}
    \label{fig:overallfigure}
\end{figure}

\textbf{Comparison across different privacy budget} As illustrated in Figure \ref{subfig:eps}, our algorithm consistently outperforms the benchmark algorithm across all privacy budget values ($\epsilon$). The improvement is especially notable in the range $0.8 \leq \epsilon \leq 2$, where the benchmark algorithm yields nearly random results, with a normalized discrepancy close to 1, while our algorithm produces results almost identical to the non-private spectral clustering.

\textbf{Comparison across different graph size} Figure \ref{subfig:n} presents a comparison with the benchmark algorithm across varying numbers of nodes ($n$). From the figure, we observe that while our algorithm performs poorly for small $n$, it achieves results identical to non-private spectral clustering when $n$ becomes sufficiently large. This aligns with our theoretical findings, which indicate that the noise introduced by our algorithm becomes negligible as the input graph size increases. 

The plot also reveals that the randomized response-based algorithm performs well only when the input graph size is small. This observation aligns with the theoretical findings of previous work \citep{mukherjee2023robustness}, which state that the required privacy budget must exceed $\Theta(\log n)$. Consequently, larger values of $n$ demand a higher privacy budget in the prior approach. In summary, our algorithm demonstrates greater precision for larger $n$, whereas the previous method performs better on very small graphs.

It is worth noting that, for the plot in Figure \ref{subfig:n} alone, we conducted the experiment on Google Colaboratory. This was necessary because our local computing environment lacked the storage capacity for the randomized response results for graphs of that size. However, we have verified that the precision results remain consistent across different computational environments.

\textbf{Comparison across different edge density} In Figures~\ref{subfig:density1} and \ref{subfig:density15}, we explore the impact of graph density by varying the probabilities \( p \) and \( q \). The experiments are conducted for all pairs \((p, q) \in \{0.05, 0.1, \ldots, 0.95\}^2\) and for $\epsilon \in \{1, 1.5\}$. Due to the large number of experiments, the graph size is reduced to 1000 for this analysis. The results show that when \( p > 0.35 \), our algorithm consistently outperforms the randomized response-based method, achieving a smaller normalized discrepancy in these cases.

When \( p \leq 0.35 \), there are instances where our algorithm performs worse than the benchmark algorithm. This occurs because the estimated minimum degree, \( \delta \), is relatively small in these cases, resulting in a larger amount of noise added in Algorithm 1. While we have theoretically shown that our algorithm can produce results comparable to original spectral clustering when \( \delta \geq \sqrt{n} \log^4 n \) (where \( n \) is the number of nodes), this analysis is valid only for large \( n \) and does not extend to cases where \( n = 1000 \). On the other hand, as shown in \citep{mohamed2022differentially}, the randomized response-based algorithm performs well when \( q \leq p \) and \( p \) is small. Consequently, in these scenarios, the randomized response method outperforms our algorithm.

We observe that when \( q > p \), the results of both algorithms deviate from those of the original spectral clustering algorithm. This outcome arises because the input graphs are not well-clustered, leading to poor performance from both the original spectral clustering method and the two algorithms in these cases.

\textbf{Computation time} Although our algorithm is designed to be executed in a distributed manner in practice, we were unable to afford the necessary computation units for handling 10,000 nodes in this experiment. As a result, all computations were performed on our server, making the computation environment different from practical scenarios. Consequently, a direct comparison of the computation times between our algorithm and the benchmark algorithm is not feasible. However, even with all computations performed on the server, the computation time for graphs with 20,000 nodes is less than 10 seconds for both algorithms, and for graphs with 1,000,000 nodes, our algorithm completes in under 1 minute. Therefore, we consider computation time to be a manageable factor for both algorithms.

\subsection{Results on Reddit graph} 
\label{subsec:reddit}

\begin{wrapfigure}{l}{0.31\textwidth}
  \vspace{-10pt}  
  \includegraphics[width=\linewidth]{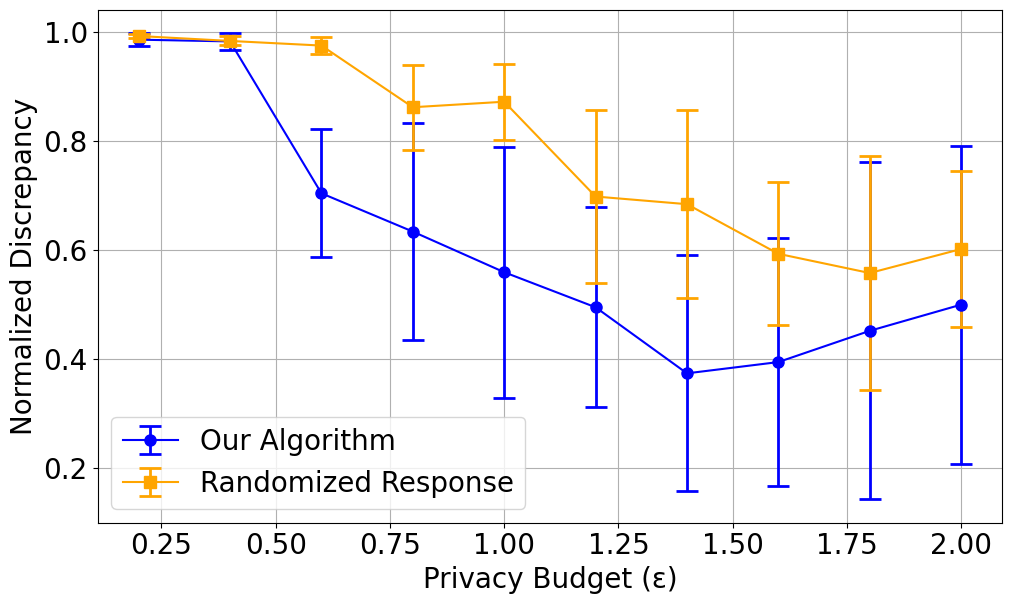}
  \caption{The normalized discrepancies of our algorithm for the graph extracted from the Reddit graph}
  \label{fig:wrapcore700}
  \vspace{-10pt}  
\end{wrapfigure}

In this section, we give our result on the real graph called Reddit graph \citep{hamilton2017inductive}. We chose this graph because it is the only large and dense networks, which are publicly available. Since the large minimum degree is required by any locally differentially private algorithm, we calculate a 700-core, 500-core, and 100-core decomposition of the graph before giving it as an input of both algorithms. Results for the 700-core decomposition graph are given in this section, while the results for the 500-core and 100-core decompositions are given in Appendix \ref{appedix:additionalReddit}.
The 700-core decomposition result contains 22,370 nodes and 31,053,002 edges.

Since the Reddit graph contains more than two clusters, we observed that $\lambda_3(B) + 1$ is very close to $\lambda_2(B) + 1$, and the value of $g$ (defined in Section \ref{subsec:assumption}) must be set as low as $1.005$. Consequently, the number of iterations required by the algorithm, calculated as $2 \log n / \log g$, increases significantly to approximately 1,700. Given that the noise size is dependent on the number of iterations, this large iteration count renders the noise size unmanageable. To address this, we limited the number of iterations to 50 for this experiment.

Our results for these graphs are presented in Figure \ref{fig:wrapcore700}. We can demonstrate that our algorithm outperforms the benchmark algorithm across all different privacy budget.
While the normalized discrepancy rapidly converges to $0$ in graphs generated by the model, it does not converge to $0$ in Figure \ref{fig:overallfigure2}. We attribute this to the Reddit graph containing more than two clusters, which results in a significant number of nodes $v_i$ with small $|\nu_i|$ (as discussed in Assumption 3 in Section \ref{subsec:assumption}). Consequently, our algorithm is unable to classify these nodes correctly.

\section{Conclusion and future work}

In this paper, we propose a locally differentially private algorithm for graph clustering that is theoretically proven to work on general graphs. Unlike most prior works, which focus on non-interactive algorithms based on randomized response, we introduce an interactive algorithm leveraging power iterative clustering. Our approach demonstrates both theoretical and experimental improvements over previous methods. By this work, we believe that interactive algorithms have the potential to become a key tool for addressing graph problems under local differential privacy.

Although our algorithm is applicable to sparse graphs, our theoretical guarantees currently hold only for dense graphs. Extending the theory to sparse graphs requires an additional condition: for any eigenvector \( \mathbf{v}_i = [v_{i,1}, \dots, v_{i,n}]^\intercal \), the ratio \( \max_{j,j'} \frac{v_{i,j}}{v_{i,j'}} \) must be small. This property, known as delocalization, has been studied in several works, such as \citep{rudelson2016no}. We plan to investigate the potential of incorporating this property into our analysis.

\section*{Acknowledgments}
Vorapong Suppakitpaisarn is supported by KAKENHI Grants JP21H05845, JP23H04377, and JP25K00369, as well as JST NEXUS Grant Number Y2024L0906031.
Sayan Mukherjee is supported by JSPS KAKENHI Grant Number 24K22830, and the Center of Innovations in Sustainable Quantum AI (JST Grant Number JPMJPF2221).

\bibliographystyle{abbrvnat}
\bibliography{neurips_2025}


\appendix

\section{Comparison with the assumptions in \citep{hehir2022consistent}}

\label{appendix:assumption}
In this section, we show that our assumptions in Section \ref{subsec:assumption} are at least as general as those in the previous work \citep{hehir2022consistent}. Specifically, any graph that yields a robust clustering result under their algorithm and analysis will also be robustly clustered by our algorithm. While their analysis is restricted to graphs generated from the stochastic block model (SBM) and the degree corrected block model (DCBM), our analysis and algorithm apply to any graph that is well-clustered and has a large minimum degree.

Equation (5.3) of \citep{hehir2022consistent} shows that when each community has size $\Omega(n)$, randomized response followed by standard spectral clustering results in a misclassification error of order $O(1)$, provided that the maximum connection probability between any two nodes is at least $\Omega(1/\sqrt{n})$. The degree distribution in such an SBM follows a binomial distribution, which can be approximated by a Gaussian distribution for large $n$. By that the minimum degree in these graphs is $\tilde{\Omega}(\sqrt{n})$ with probability at least $1 - 1/n$. Therefore, the SBM and DCBM considered in \citep{hehir2022consistent} satisfies Assumption (1).

Regarding Assumption (2), we can infer from \citep{JMLR:v22:20-391} that when the number of nodes $n$ is large, $p$ denotes the probability that two nodes within the same cluster are connected, and $q$ denotes the probability that two nodes from different clusters are connected, then the ratio 
$g = \frac{\lambda_2(B) + 1}{\lambda_3(B) + 1}$
is at least $\frac{2p}{p + q}$. Since typically $p \gg q$, this implies $g \approx 2$, indicating that the graphs generated from SBM satisfies the assumption. Empirically, any well-clustered graph---including those generated by the DCBM or similar models---tends to have a large $g$, and thus also satisfies this assumption.

Any well-clustered graph in which each cluster has size in 
 satisfies the conditions of Proposition \ref{prop:app_eigenvector_components}. Therefore, graphs generated from the SBM and other clustered models satisfy Assumption (3).

\section{Eigenvector components}
\label{appendix:eigenvector}

In this section, we analyze the Laplacian matrix of the graph \( G \), defined as \( L = I - B \). For each \( i \), let \( \lambda_i(L) = 1 - \lambda_i(B) \). It follows that \( \lambda_i(L) \) is an eigenvalue of \( L \), and the eigenvalues are ordered as \( \lambda_1(L) \leq \dots \leq \lambda_n(L) \). Moreover, the eigenvector \( \mathbf{v}_i(B) \) associated with \( \lambda_i(B) \) is also an eigenvector of \( L \) corresponding to \( \lambda_i(L) \). For simplicity, throughout this section, we denote \( \lambda_i(L) \) by \( \lambda_i \) and \( \mathbf{v}_i(B) \) by \( \mathbf{v}_i = [v_{i,1}, \dots, v_{i,n}]^\intercal \).


\begin{proposition}
	\label{prop:app_eigenvector_components}
    Assume that

    (i) Let $V(G)=A\sqcup B$ be a bipartition of $G$ with $v_{2,j} \ge 0$ for $v_j \in A$, $v_{2,j}\le 0$ for $v_j \in B$.
    Then, the cut $(A,B)$ has conductance $\phi$ satisfying $\phi/\lambda_3\le 0.12$.
    
    (ii) Let $\epsilon$ and $c$ be a constant. For a subset $S\subseteq V$ and vertex $v_j\in S$, let us call $v_j$ to be $(\epsilon, S)$-average if $d_j\ge \epsilon d(S)$, where $d(S)=\Vol(S)/|S|$ is the average degree of the vertices in $S$.
    Let $A_\epsilon$ and $B_\epsilon$ denote the set of $(\epsilon,A)$-average nodes of $A$ and $(\epsilon,B)$-average nodes of $B$, respectively.
    Assume that $|A_\epsilon|\ge c|A|$ and $|B_\epsilon|\ge c|B|$.

    Then,
    \begin{equation}
    \label{eq:app_prop_eigenvector_components_1}
        |v_{2,j}|\ge \left\{\begin{array}{cl}\frac{\epsilon^{1/2}c}4\cdot \sqrt{\frac{d_j}{nd(A)}} - 2\sqrt{\frac{\phi}{\lambda_3}}, & v\in A\\
        \frac{\epsilon^{1/2}c}4\cdot \sqrt{\frac{d_j}{nd(B)}} - 2\sqrt{\frac{\phi}{\lambda_3}}, & v\in B
        \end{array}\right.
    \end{equation}
    
    Consequently, for $v_j\in A_\epsilon\cup B_\epsilon$, which is at least $c$ fraction of the vertices of $G$, we have
    \begin{equation}
    \label{eq:app_prop_eigenvector_components_2}
    	|v_{2,j}|\ge \frac{\epsilon c}{4}\cdot \frac1{\sqrt n} - 2\sqrt{\frac{\phi}{\lambda_3}}.
    \end{equation}
\end{proposition}
\begin{proof}
    Let us define the normalized indicator variables
    \[g_A(j)=\left\{\begin{array}{cl}\frac{d_j^{1/2}}{\Vol(A)^{1/2}}, & v_j\in A\\ 0, & v_j \in B \end{array}\right.\text{ and } g_B(j)=\left\{\begin{array}{cl} 0, & v_j \in A\\\frac{d_j^{1/2}}{\Vol(B)^{1/2}}, & v_j\in B\end{array}\right..\]
    Let the vector $g_A = [g_A(1), \dots, g_A(n)]^\intercal$, $g_B = [g_B(1), \dots, g_B(n)]^\intercal$, and, for any vector $\mathbf{v}$, the Rayleign quotient of $\mathbf{v} = [x_1, \dots, x_n]^\intercal$, denoted by $\mathcal{R}(\mathbf{v})$, is $\frac{\mathbf{v}^\intercal L \mathbf{v}}{\mathbf{v}^\intercal \mathbf{v}}$. We show the following regarding the Rayleigh quotients $\mathcal R_L(g_A)$ and $\mathcal R_L(g_B)$.
    \begin{claim}
    \label{clm:app_Rayliegh_indicators}
        $\phi \ge \max\{\mathcal R_L(g_A), \mathcal R_L(g_B)\}$. 
    \end{claim}
    \textit{Proof of Claim~\ref{clm:app_Rayliegh_indicators}.}
    Observe that the Rayleigh quotient of $L$ satisfies,
    \begin{equation}
    \label{eq:app_Rayleigh_Laplacian}
        \mathcal R_L(\mathbf{v}) 
        = \frac{\mathbf{v}^\intercal L\mathbf{v}}{\mathbf{v}^\intercal \mathbf{v}}
        = 1 - \frac{\sum_{i=1}^n\sum_{j=1}^n \frac{a_{ij}}{d_i}x_ix_j}{\sum_{i=1}^n x_i^2} 
        = 1 - \frac{\sum_{\{i,j\}\in E} \left(\frac1{d_i}+\frac1{d_j}\right)x_ix_j}{\sum_{i=1}^nx_i^2}.
    \end{equation}
    
    Since $\|g_A\|^2=1$, we have
    \[
    \begin{aligned}
        \mathcal R_L(g_A)=1-\sum_{\{i,j\}\in E}\left(\frac1{d_i}+\frac1{d_j}\right)g_A(i)g_A(j)&=1-\sum_{\{i,j\}\in E(A)} \left(\frac1{d_i}+\frac1{d_j}\right)\cdot \frac{\sqrt{d_id_j}}{\Vol(A)}\\
        &\le 1 - \sum_{\{i,j\}\in E(A)}\frac{2}{\Vol(A)} = \frac{\Vol(A) - 2e(A)}{\Vol(A)}\\
        & = \frac{e(A,B)}{\Vol(A)} \le \phi.
    \end{aligned}
    \]
    Similarly, we have $\mathcal R_L(g_B)\le \phi$, completing the proof of Claim~\ref{clm:app_Rayliegh_indicators}.\hfill{$\blacksquare$}

    For the rest of the proof, let us denote $t:=\phi/\lambda_3$. Recall that $\mathbf{v}_1 = [1/\sqrt n, \ldots, 1/\sqrt n]^\intercal$.
    We will make use of the following lemmas from the structure theorem (Theorem 3.1) of \citep{peng2015partitioning}, but with a different notation and error estimates.
    \begin{lemma}
    \label{lem:app_PengLem1}
        Let $\hat g_A$, $\hat g_B$ be the projections of $g_A$, $g_B$ onto the space spanned by the first two eigenvectors $\{\mathbf{v}_1, \mathbf{v}_2\}$ of $L$.
        Then, 
        \begin{equation}
            \max\{\|\hat g_A-g_A\|^2, \|\hat g_B-g_B\|^2\}\le t.
        \end{equation}
    \end{lemma}
    \textit{Proof of Lemma~\ref{lem:app_PengLem1}.}
    Let $\mathbf{v}_3,\ldots, \mathbf{v}_n$ be normalized eigenvectors of $\lambda_3,\ldots,\lambda_n$ of $L$.
    Say $g_A = \alpha_1 \mathbf{v}_1 + \cdots + \alpha_n \mathbf{v}_n$ and $g_B=\beta_1\mathbf{v}_1+\cdots+\beta_n\mathbf{v}_n$ are representations of $g_A$ and $g_B$ in the $L$-eigenbasis.
    Clearly $\hat g_A=\alpha_1\mathbf{v}_1+\alpha_2\mathbf{v}_2$ and $\hat g_B = \beta_1\mathbf{v}_1+\beta_2\mathbf{v}_2$.
    Then, note that as $\mathbf{v}^\intercal_i\mathbf{v}_j=0$ for every $i\neq j$, 
    \[\mathcal R_L(g_A) = \sum_{i=1}^n\alpha_i\mathbf{v}_i^\intercal\cdot L \cdot \sum_{i=1}^n\alpha_i\mathbf{v}_i=\sum_{i=1}^n \alpha_i^2 \mathbf{v}_i^\intercal L\mathbf{v}_i = \sum_{i=1}^n\alpha_i^2\lambda_i.\]
    But $\lambda_1=0$, leading us to $\mathcal R_L(g_A)\ge \alpha_2^2\lambda_2 + (\alpha_3^2+\cdots+\alpha_n^2)\lambda_3=\alpha_2^2\lambda_2 + \|\hat g_A-g_A\|^2\lambda_3\ge \|\hat g_A-g_A\|^2\lambda_3$.
    Thus, $\|\hat g_A-g_A\|^2\le \mathcal R_L(g_A)/\lambda_3 \le \phi/\lambda_3$ by Claim~\ref{clm:app_Rayliegh_indicators}.
    The proof for $\|\hat g_B-g_B\|^2$ is exactly analogous.\hfill{$\blacksquare$}

    One of the main ideas used in  \citep{peng2015partitioning} is that if $\hat g_A$ and $\hat g_B$ are independent, then $\text{Span}(\{\mathbf{v}_1, \mathbf{v}_2\})=\text{Span}(\{\hat g_A, \hat g_B\})$, implying that $\mathbf{v}_1$ and $\mathbf{v}_2$ can be written as linear combinations of the projected indicator vectors $\hat g_A$ and $\hat g_B$, say $\mathbf{v}_2 = \eta_1\hat g_A+\eta_2\hat g_B$, implying that $\|\mathbf{v}_2-\eta_1g_A-\eta_2g_B\|$ is small.

    Let us now continue with the argument.
    \begin{claim}
        \label{clm:app_LinearInd_indicatorprojections}
        $\hat g_A$ and $\hat g_B$ are linearly independent.
    \end{claim}
    \textit{Proof of Claim~\ref{clm:app_LinearInd_indicatorprojections}}.
    By Lemma~\ref{lem:app_PengLem1}, we have $\|\hat g_A\|^2\ge 1-t$ and $\|\hat g_B\|^2\ge 1-t$.
    On the other hand,
    \begin{equation}
    \label{eq:app_hat_gA-hat_gB-dotproduct}
    \begin{aligned}
        |\langle \hat g_A,\hat g_B\rangle| &= |\langle \hat g_A-g_A+g_A,\hat g_B-g_B+g_B\rangle| \\
        &\le |\langle \hat g_A-g_A, \hat g_B-g_B\rangle| + |\langle g_A,\hat g_B-g_B\rangle| + |\langle \hat g_A-g_A, g_B\rangle| \\
        & \le \|\hat g_A-g_A\|\|\hat g_B-g_B\|+\|\hat g_A-g_A\|+\|\hat g_B-g_B\|\\
        & \le t + 2\sqrt{t}.
    \end{aligned}
    \end{equation}
    Since $t\le0.12<\frac12(2-\sqrt3)$, we have $t+2\sqrt t < 1-t$, implying $|\langle \hat g_A, \hat g_B\rangle| < \|\hat g_A\|\|\hat g_B\|$.
    As this implies a strict inequality in the Cauchy-Schwarz inequality, we have $\hat g_A\nparallel \hat g_B$.\hfill{$\blacksquare$}

    As discussed earlier, Claim~\ref{clm:app_LinearInd_indicatorprojections} implies that there exist $\eta_1,\eta_2\in\mathbb R$ such that $\mathbf{v}_2 = \eta_1\hat g_A + \eta_2\hat g_B$.
    Suppose $\mathbf{v}'_2 = \eta_1g_A+\eta_2g_B$, and $\eta=\|\mathbf{v}'_2\|=\sqrt{\eta_1^2+\eta_2^2}$.
    Note that, using (\ref{eq:app_hat_gA-hat_gB-dotproduct}),
    \[
    \begin{aligned}
    1 = \|\mathbf{v}_2\|^2 & \ge \eta_1^2\|\hat g_A\|^2 + \eta_2^2\|\hat g_B\|^2 - 2|\eta_1\eta_2\langle \hat g_A, \hat g_B\rangle| \\
    & \ge \eta_1^2(1-t) + \eta_2^2(1-t) - (\eta_1^2+\eta_2^2) (t+2\sqrt t)\\
    & = \eta^2(1-2t-2\sqrt{t}).
    \end{aligned}
    \]
    Moreover, since $t\le 0.12$, we have
    \begin{equation}
        \label{eq:app_eta_bounds}
        \eta^2 \le \frac{1}{1-2t-2\sqrt t} < 16.
    \end{equation}
    Moreover, by the triangle inequality and Cauchy-Schwarz inequality, 
    \begin{equation}
    \begin{aligned}
        \|\mathbf{v}_2-\mathbf{v}'_2\|^2 &= \|\eta_1(\hat g_A-g_A)+\eta_2(\hat g_B-g_B)\|^2 \\
        &\le \left(|\eta_1|\sqrt t+|\eta_2|\sqrt t\right)^2\\
        & \le 2t\eta^2
    \end{aligned}
    \end{equation}
    Therefore, we have that
    \[
    2\eta \langle \mathbf{v}_2, \textstyle\frac1\eta \mathbf{v}'_2\rangle=2\langle \mathbf{v}_2,\mathbf{v}'_2\rangle= \|\mathbf{v}_2\|^2+\|\mathbf{v}'_2\|^2-\|\mathbf{v}_2-\mathbf{v}_2'\|^2\ge 1+\eta^2-2t\eta^2,
    \]
    leading us to
    \begin{equation}
    \label{eq:app_projection_onto_x2}
        \langle \mathbf{v}_2, \textstyle\frac1\eta \mathbf{v}'_2\rangle\ge \frac{1+\eta^2}{2\eta} - \eta t \ge 1 - \eta t.
    \end{equation}
    
    Basically, this means that $\mathbf{v}_2$ is closely aligned with the normalized vector $\frac1\eta \mathbf{v}'_2$.
    We now show a lemma that relates the components of two such vectors.

    \begin{lemma}
        \label{lem:app_componentsEig}
        Let $\mathbf{v} = [u_1, \dots, u_n]^\intercal$ be a unit eigenvector of $L$ and $\mathbf{v}' = [u'_1, \dots, u'_n]^\intercal$ be any unit vector with $\langle \mathbf{v}, \mathbf{v}'\rangle \ge 1-\epsilon^2$ for some $\epsilon>0$.
        Then, for each $1\le j\le n$, we have
        \[
        |u'_j|\le |u_j|+\epsilon.
        \]
    \end{lemma}
    \textit{Proof of Lemma~\ref{lem:app_componentsEig}.}
    Let $\{\mathbf{v},\mathbf{z}_1,\ldots \mathbf{z}_{n-1}\}$ be a orthonormal basis of eigenvectors of $L$, and, for all $i$, let $\mathbf{z}_i = [z_{i,1}, \dots, z_{i,n}]^\intercal$.
    Since $\mathbf{v}'=\langle \mathbf{v}, \mathbf{v}'\rangle\cdot \mathbf{v} + \sum_{i=1}^{n-1}\langle \mathbf{v}', \mathbf{z}_i\rangle \cdot \mathbf{z}_i$, this implies that for any $1\le j\le n$,
    \[
    \begin{aligned}
        |u'_j| &\le |\langle \mathbf{v},\mathbf{v}'\rangle| |u_j| + \sum_{i=1}^{n-1}|\langle \mathbf{v}', \mathbf{z}_i\rangle| |z_{i,j}|\\
        &\le |u_j| + \left(\sum_{i=1}^{n-1}\langle \mathbf{v}', \mathbf{z}_i\rangle^2\right)^{1/2}\left(\sum_{i=1}^{n-1}z_{i,j}^2\right)^{1/2}\\
        & \le |u_j|+\epsilon,
    \end{aligned}
    \]
    where the last step follows from the fact that $\sum_{i=1}^{n-1}\langle \mathbf{v}',\mathbf{z}_i\rangle^2+\langle \mathbf{v}',\mathbf{v}\rangle^2=\|\mathbf{v}'\|^2=1$, and $\sum_{i=1}^{n-1}z_{i,j}^2+u_j^2=1$.\hfill{$\blacksquare$}

    Hence, by virtue of Lemma~\ref{lem:app_componentsEig}, (\ref{eq:app_eta_bounds}) and (\ref{eq:app_projection_onto_x2}), we obtain
    \begin{equation}
    \label{eq:app_abs_x2_v_geq_midway}
        |v_{2,j}| \ge \frac1\eta|v'_{2,j}|-\sqrt{\eta t} = \frac 1\eta |\eta_1g_A(j)+\eta_2g_B(j)| - \sqrt{\eta t} = 
        \left\{\begin{array}{cl}\frac{|\eta_1|}\eta\cdot \frac{d_j^{1/2}}{\Vol(A)^{1/2}} - \sqrt{\eta t}, & v_j\in A \\
        \frac{|\eta_2|}\eta\cdot \frac{d_j^{1/2}}{\Vol(B)^{1/2}} - \sqrt{\eta t}, & v_j\in B\end{array}\right.
    \end{equation}
    
    Finally, we need to show that $\min\{|\eta_1|,|\eta_2|\}\ge \epsilon^{1/2}c$.
    For this part of the proof, we shall use the assumption (ii) of our proposition.

    \begin{claim}
    \label{clm:app_eta_i_big}
        $|\eta_1|\ge c\cdot \left(\frac{\epsilon|A|}n\right)^{1/2}$ and $|\eta_2|\ge c\cdot \left(\frac{\epsilon|B|}n\right)^{1/2}$.
    \end{claim}
    \textit{Proof of Claim~\ref{clm:app_eta_i_big}.}
    Recall from the proof of Lemma~\ref{lem:app_PengLem1} that $\hat g_A=\alpha_1\mathbf{v}_1+\alpha_2\mathbf{v}_2$ and $\hat g_B=\beta_1\mathbf{v}_1+\beta_2\mathbf{v}_2$.
    These equations, along with $\mathbf{v}_2=\eta_1\hat g_A+\eta_2\hat g_B$, allow us to solve exactly for $\eta_1$ and $\eta_2$ as,
    \[\eta_1=\frac{\beta_1}{\alpha_2\beta_1-\alpha_1\beta_2}\mbox{ and }\eta_2=\frac{-\alpha_1}{\alpha_2\beta_1-\alpha_1\beta_2}.\]
    First, we note that $|\alpha_2\beta_1-\alpha_1\beta_2|\le(\alpha_1^2+\alpha_2^2)^{1/2}(\beta_1^2+\beta_2^2)^{1/2}\le \|g_A\|\|g_B\|=1$, so it suffices to lower bound $|\alpha_1|$ and $|\beta_1|$.
    We have that:
    \[
    \begin{aligned}
    |\alpha_1|=|\langle g_A, \mathbf{v}_1\rangle| = \frac{1}{\sqrt n}\sum_{v_j \in A}\frac{d_j^{1/2}}{\Vol(A)^{1/2}} &\geq \frac1{\sqrt n}\sum_{v_j \in A_\epsilon}\frac{d_j^{1/2}}{(|A|d(A))^{1/2}}\\
    &\ge \frac1{\sqrt n}\cdot |A_\epsilon|\cdot \left(\frac{\epsilon}{|A|}\right)^{1/2}\\
    &\ge c\cdot \left(\frac{\epsilon|A|}n\right)^{1/2}.
    \end{aligned}
    \]
    By a similar argument, we have $|\beta_1|\ge c\cdot \left(\frac{\epsilon|B|}n\right)^{1/2}$, finishing the proof of Claim~\ref{clm:app_eta_i_big}.
    \hfill{$\blacksquare$}
    
    Claim~\ref{clm:app_eta_i_big}, (\ref{eq:app_abs_x2_v_geq_midway}) and $\eta\le 4$ leads us to, for $v_j \in A$,
    \[
    |v_{2,j}|\ge \frac{c}4\cdot \frac{\epsilon^{1/2}|A|^{1/2}}{n^{1/2}}\cdot \frac{d_j^{1/2}}{\Vol(A)^{1/2}}-2\sqrt t = \frac{c\epsilon^{1/2}}{4}\cdot \sqrt{\frac{d_j}{nd(A)}} - 2\sqrt t,
    \]
    which proves the inequality (\ref{eq:app_prop_eigenvector_components_1}) for $v_j \in A$.
    The argument for $v_j \in B$ is analogous.
    
    Finally, the inequality (\ref{eq:app_prop_eigenvector_components_2}) directly follows (\ref{eq:app_prop_eigenvector_components_1}) via the definitions of $A_\epsilon$ and $B_\epsilon$.
    
\end{proof}

\section{Elimination of the leading eigenvector} 
\label{appendix:elimination}

The following proposition shows that the third term in the calculation at Line 6 of Algorithm \ref{alg:private_modified_PIC} eliminates the leading eigenvector of \( W \). Consequently, the leading eigenvector of \(\tilde{W}\) becomes the second eigenvector of \( W \).

\begin{proposition}
\label{lem:delete_largest_eigval}
    Let $W=\frac12(I+D^{-1}A)$ be the lazy random walk matrix for a graph on $n$ vertices.
    Let $J = (\mathsf{j}_{i,j})_{1 \leq i,j \leq n}$ be a matrix such that $\mathsf{j}_{i,j} = 1$ for all $i,j$. Define $\tilde{W}=W-\frac1nJ$.
    Then, for all \(i \geq 1\), \(\lambda_i(\tilde{W}) = \lambda_{i+1}(W)\) and \(\mathbf{v}_n(\tilde{W}) = \mathbf{v}_1(W)\). Additionally, \(\mathbf{v}_n(\tilde{W}) = \mathbf{v}_1(W) = [\frac1{\sqrt{n}}, \dots, \frac1{\sqrt{n}}]^\intercal\) and \(\lambda_n(\tilde{W}) = 0\).
\end{proposition}
\begin{proof}
Recall that \(\mathbf{v}_1(W) = \left[ \frac{1}{\sqrt{n}}, \dots, \frac{1}{\sqrt{n}} \right]^\intercal\) and \(\lambda_1(W) = 1\). We have:
\begin{eqnarray*}
\tilde{W} \cdot \mathbf{v}_1(W) & = & W \cdot \mathbf{v}_1(W) - \frac{1}{n} J \mathbf{v}_1(W)  = \mathbf{v}_1(W) - \left[ \frac{1}{\sqrt{n}}, \dots, \frac{1}{\sqrt{n}} \right]^\intercal = \mathbf{0}.    
\end{eqnarray*}

Therefore, the vector \(\left[ \frac{1}{\sqrt{n}}, \dots, \frac{1}{\sqrt{n}} \right]^\intercal\) is an eigenvector of \(\tilde{W}\) with eigenvalue \(0\). Since \(0\) is the minimum eigenvalue of \(\tilde{W}\), it follows that \(\mathbf{v}_n(\tilde{W}) = \mathbf{v}_1(W)\) and \(\lambda_n(\tilde{W}) = 0\).

Next, let us consider $\mathbf{v}_i(W)$ for $i \geq 2$. Since, $\mathbf{v}_i(W) \perp \mathbf{v}_1(W)$, we obtain that the sum of all elements in $\mathbf{v}_i(W)$ is zero.
Thus,
\[
\tilde{W} \mathbf{v}_i(W) = W \mathbf{v}_i(W) - \frac{1}{n} J \mathbf{v}_i(W) = \lambda_i(W) \mathbf{v}_i(W).
\]
This implies that, for all \(i \geq 2\), \(\mathbf{v}_i(W)\) is also an eigenvector of \(\tilde{W}\) with the same eigenvalue. Consequently, as the largest eigenvalue of \(W\) becomes the smallest eigenvalue of \(\tilde{W}\), we have \(\lambda_{i - 1}(\tilde{W}) = \lambda_i(W)\) and \(\mathbf{v}_{i - 1}(\tilde{W}) = \mathbf{v}_i(W)\).
\end{proof}

\section{Privacy}
\label{sec:privacy}
The following theorem discuss our algorithm's privacy. \begin{theorem}
    Algorithm \ref{alg:private_modified_PIC} is $\epsilon$-edge LDP.
\end{theorem}
\begin{proof}
We perform \(T + 1\) edge-local Laplacian queries to all users: one at Line 1 and \(T\) queries at Line 6. At Line 1, the degree \(d_i\) has a sensitivity of one. Since the Laplace noise is set to \(10/\epsilon\), the privacy budget for the publication at Line 1 is \(\epsilon / 10\).

When any \(a_{i,j}\) is changed, the value of \(x_i^{(t)}\) calculated at Line 6 changes by at most \(\frac{1}{2} \max_j \frac{|x_j^{(t-1)}|}{d_j}\). Therefore, the sensitivity of the publication at Line 6 is \(\frac{1}{2} \max_j \frac{|x_j^{(t-1)}|}{d_j} \leq \frac{1}{2} \max_j \frac{|x_j^{(t-1)}|}{\delta}\). The privacy budget for each publication at Line 6 is \(\frac{9}{10} \cdot \frac{\epsilon}{T}\).
Since there are \(T\) publications at Line 6, the total privacy budget of Algorithm \ref{alg:private_modified_PIC} is \(\frac{\epsilon}{10} + T \cdot \frac{9}{10} \cdot \frac{\epsilon}{T} = \epsilon\).
\end{proof}

\section{Minimum degree estimation}
\label{appendix:minimumDegree}
We will now demonstrate that the value of \(\delta\) computed in Line 2 of Algorithm \ref{alg:private_modified_PIC} has a low probability of overestimating the minimum degree of the input graph. This implies that, with large probability, we do not need to modify the input graph in Line 3 of the algorithm.
\begin{proposition}
    With probability at least $1 - \zeta$, we have $\delta < \min_i d_i$.  
\end{proposition}
\begin{proof}
    We have $\delta > \min_i d_i$ only if there is $\tilde{d}_i$ such that $\tilde{d}_i - \frac{10}{\epsilon} \log \frac{n}{2\zeta} > d_i$. This implies that the value sampled from the Laplace distribution at Line 1, denoted by $\mathsf{l}_i$ is larger than $\frac{10}{\epsilon} \log \frac{n}{2\zeta}$. By the property of the Laplace distribution, for all $i$, we have that:
    \begin{eqnarray*}
        \Pr\left[\mathsf{l}_i > \frac{10}{\epsilon} \log \frac{n}{2\zeta}\right] & = & \frac{1}{2} \exp\left( - \frac{10}{\epsilon} \log \frac{n}{2\zeta} / \frac{10}{\epsilon} \right)  = \zeta / n.
    \end{eqnarray*}
    Then, by the union bound, the probability that there is an index $i$ such that $\mathsf{l}_i > \frac{10}{\epsilon} \log \frac{n}{2\zeta}$ is not greater than $\zeta$.
\end{proof}

Suppose that $\zeta = \frac{1}{n}$.
In the next proposition, we shown that $\delta \geq \sqrt{n} \log^4 n$ with large probability.
\begin{proposition}
    Assuming that the input graph $G$ satisfies Assumption~(1), i.e., its minimum degree is at least $2\sqrt{n} \log^4 n$, it holds that
\[
\Pr[\delta < \sqrt{n} \log^4 n] \leq \frac{1}{2n}.
\] \label{prop:delta}
\end{proposition}
\begin{proof}
   In Line 2 of Algorithm \ref{alg:private_modified_PIC}, Laplacian noise with a scale of \(\frac{10}{\epsilon}\) is added. It follows that \(\tilde{d}_i < d_i - \frac{20}{\epsilon} \log n\) if the noise added to \(d_i\) is less than \(-\frac{20}{\epsilon}\). This event occurs with probability 
\[
\frac{1}{2} \exp\left(-\frac{20/\epsilon \cdot \log n}{10/\epsilon}\right) = \frac{1}{2n^2}.
\]

Using the union bound, we have:
\[
\Pr\left[\min_i \tilde{d}_i < \min_i d_i - \frac{20}{\epsilon} \log n \right] \leq \Pr\left[\tilde{d}_i < d_i - \frac{20}{\epsilon} \log n \text{ for some } i \right] \leq \frac{1}{2n}.
\]

Given that \(\delta = \min_i \tilde{d}_i - \frac{10}{\epsilon} \log \frac{n}{2\zeta}\), and under the assumption in Section \ref{subsec:assumption} that the minimum degree of the network is at least \(2 \sqrt{n} \log^4 n\), we can bound:
\[
\Pr\left[\delta < \sqrt{n} \log^4 n \right] \leq \Pr\left[\delta < \min_i d_i - \frac{20}{\epsilon} \log n - \frac{10}{\epsilon} \log \frac{n}{2\zeta}\right] \leq \frac{1}{2n},
\]
for sufficiently large \(n\).
\end{proof}

\section{Size of Laplace noise}
\label{appendix:sizeofnoise}
In this section, we analyze the effect of adding the Laplace noise at Line 6 of the algorithm. Let the noise added by the node $i$ at the iteration $t$ is $y_{i}^{(t)}$. Define the vector $\mathbf{y}^{(t)}$ as $[y^{(t)}_1, \dots, y^{(t)}_n]^\intercal$. Also, for all $i,t$, let $e_{i}^{(t)}$ be a real number such that $\mathbf{y}^{(t)} = e_1^{(t)} \mathbf{v}_1(\tilde{W}) + \dots + e_n^{(t)} \mathbf{v}_n(\tilde{W})$.

Let the initial vector denoted by $\mathbf{x}^{(0)} = c_1 \mathbf{v}_1(\tilde{W}) + \dots + c_n \mathbf{v}_n(\tilde{W})$, and the final vector is denoted by $\mathbf{x}^{(T)}$. We obtain the following lemma by the notation.
\begin{lemma}
    Let $\tilde{c}_1, \dots, \tilde{c}_n$ be numbers such that $\mathbf{x}^{(T)} = \tilde{c}_1 \mathbf{v}_1(\tilde{W}) + \dots + \tilde{c}_n \mathbf{v}_n(\tilde{W})$. We obtain that $\tilde{c}_i = c_i\lambda_i(\tilde{W})^T + e_i^{(1)} \lambda_i(\tilde{W})^{T - 1} + \dots + e_i^{(T)}$. \label{lem:e} 
\end{lemma}
\begin{proof}
    To prove the statement, let \(\mathsf{c}_i^{(t)} = c_i\lambda_i(\tilde{W})^t + e_i^{(1)} \lambda_i(\tilde{W})^{t - 1} + \dots + e_i^{(t)}\). We proceed by induction on \(t\) to show that, for all \(t \geq 0\), \(\mathbf{x}^{(t)} = \mathsf{c}^{(t)}_1 \mathbf{v}_1(\tilde{W}) + \dots + \mathsf{c}^{(t)}_n \mathbf{v}_n(\tilde{W})\).
When \(t = 0\), \(\mathsf{c}_i^{(0)} = c_i\), so the statement holds directly by the definition of the notation.
Assume the statement is true for \(t - 1\); that is, \(\mathbf{x}^{(t - 1)} = \mathsf{c}^{(t - 1)}_1 \mathbf{v}_1(\tilde{W}) + \dots + \mathsf{c}^{(t - 1)}_n \mathbf{v}_n(\tilde{W})\). Then, for \(\mathbf{x}^{(t)}\), we have
\[
\mathbf{x}^{(t)} = \tilde{W} \cdot \mathbf{x}^{(t - 1)} + \mathbf{y}^{(t)}.
\]
Expanding this using the induction hypothesis gives
$$\mathbf{x}^{(t)} = (\mathsf{c}^{(t - 1)}_1 \lambda_1(\tilde{W}) + e_1^{(t)}) \mathbf{v}_1(\tilde{W}) + \dots + (\mathsf{c}^{(t - 1)}_n \lambda_n(\tilde{W}) + e_n^{(t)}) \mathbf{v}_n(\tilde{W}).$$
Thus, we obtain \(\mathbf{x}^{(t)} = \mathsf{c}^{(t)}_1 \mathbf{v}_1(\tilde{W}) + \dots + \mathsf{c}^{(t)}_n \mathbf{v}_n(\tilde{W})\), completing the induction.
\end{proof}
From now, let $\mathbf{v}_i(\tilde{W}) = [v_{i,1}, \dots, v_{i,n}]^\intercal$. We will now calculate the size of each variable. Recall from Line 4 of Algorithm \ref{alg:private_modified_PIC} that $x_i^{(0)}$ is sampled from the Gaussian distribution with expected value $0$ and standard deviation $1$.
\begin{lemma}
   For each \( i \), the variable \( c_i \) is a normal random variable with mean \( 0 \) and standard deviation \( 1 \). Furthermore, for $i \neq j$, $c_i$ is independent to $c_j$.\label{lem:c}
\end{lemma}
\begin{proof}
    Since, for all $i$, the eigenvector $\mathbf{v}_i(\tilde{W})$ is a unit vector and $c_i=\langle\mathbf{x}^{(0)},\mathbf{v}_i(\tilde{W})\rangle$, we have that $c_i = \sum_j v_{i,j} x_j^{(0)}$. Because $c_i$ is a linear combination of normal random variables, $c_i$ is a normal random variable. Furthermore, 
    $$\mathbb{E}[c_i] = v_{i,1} \mathbb{E}[x^{(0)}_1] + \cdots + v_{i,n} \mathbb{E}[x^{(0)}_n] = 0,$$
    and
        $$\mathrm{Var}(c_i) = v_{i,1}^2 \mathrm{Var}[x^{(0)}_1] + \cdots + v_{i,n}^2 \mathrm{Var}[x^{(0)}_n] = v_{i,1}^2 + \cdots v_{i,n}^2 = 1.$$ 

    Since \(\mathbf{v}_i(\tilde{W})\) is orthogonal to \(\mathbf{v}_j(\tilde{W}) \) for \(i \neq j\), the coefficients \(c_i\) and \(c_j\), which are the dot products of \(\mathbf{x}^{(0)}\) with \(\mathbf{v}_i(\tilde{W})\) and \(\mathbf{v}_j(\tilde{W})\) respectively, are independent of each other.
\end{proof}
Next, we give analyze the variables $e_i^{(t)}$. We observe that, although the random variable is a linear combination of the Laplace variables \( y_j^{(t)} \), it is not itself Laplace-distributed.
\begin{lemma} For all $t$ and $i$, we have $\mathbb{E}[e_i^{(t)}] = 0$.
    Furthermore, for all $t$ and all $i \neq j$, $\mathrm{Cov}(e_i^{(t)}, e_j^{(t)}) = 0$.
    \label{lem:eprop}
\end{lemma}
\begin{proof}
According to Line 6 of Algorithm \ref{alg:private_modified_PIC}, for all \( t \) and \( i \neq j \), the variables \( y_i^{(t)} \) and \( y_j^{(t)} \) are independent, with \( \mathbb{E}(y_i^{(t)}) = \mathbb{E}(y_j^{(t)}) = 0 \) and \( \mathrm{Var}(y_i^{(t)}) = \mathrm{Var}(y_j^{(t)}) \). The variable \( e_i^{(t)} \) is defined as the dot product between \( \mathbf{v}_i(\tilde W) \) and \( \mathbf{y}^{(t)} \). Specifically, if \( \mathbf{v}_i(\tilde{W}) = [v_{i,1}, \dots, v_{i,n}]^\intercal \), then \( e_i^{(t)} = \sum_j v_{i,j} y_j^{(t)} \). Consequently, \( \mathbb{E}(e_i^{(t)}) = \sum_j v_{i,j} \mathbb{E}[y_j^{(t)}] = 0 \). 

Next, for \( i \neq j \), we examine the covariance between \( e_i^{(t)} \) and \( e_j^{(t)} \), denoted as \( \mathrm{Cov}(e_i^{(t)}, e_j^{(t)}) \). Since \( \mathbb{E}(e_i^{(t)}) = \mathbb{E}(e_j^{(t)}) = 0 \), $\{y_1^{(t)},\ldots,y_n^{(t)}\}$ are independent with mean $0$, and \( \mathbf{v}_i \) is orthogonal to \( \mathbf{v}_j \), we have:
    \begin{align*}
        \mathrm{Cov}(e_i^{(t)}, e_j^{(t)})  &=  \mathbb{E}\left[\sum_{i',j'} v_{i,i'} y_{i'}^{(t)} v_{j,j'} y_{j'}^{(t)}\right] 
         \\ &=  \sum_{i',j' } v_{i,i'} v_{j,j'} \mathbb{E}[y_{i'}^{(t)}y_{j'}^{(t)}] \\ &=\sum_{k} v_{i,k}v_{j,k}\mathbb{E}[(y_k^{(t)})^2] \\&= \mathbb E[(y_1^{(t)})^2]\cdot\sum_{k} v_{i,k}v_{j,k}\\ &= 0.
    \end{align*}
\end{proof}
Let \( C_t \) represent the scale of the Laplace noise in Line 6 during the \( t \)-th iteration of Algorithm \ref{alg:private_modified_PIC}. By definition, $\mathrm{Var}(y_i^{(t)})=2C_t^2$ for every $i$. The variance of $e_i^{(t)}$ is discussed in the following lemma. Our proof draws on ideas from the paper \citep{li2023tail}.
\begin{lemma}
    For all $i$ and $t$, the variance of $e_i^{(t)}$ is $2 \cdot C_t^2$. Furthermore, $\Pr[e_i^{(t)} \geq \sqrt{2} C_t \log n] \leq \frac{e}{n}$.
    \label{lem:eprop2}
\end{lemma}
\begin{proof}
Based on the argument in the proof of Lemma \ref{lem:eprop}, we have \( e_i^{(t)} = \sum_j v_{i,j} y_j^{(t)} \). Consequently, \( \mathrm{Var}(e_i^{(t)}) = \sum_j v_{i,j}^2 \mathrm{Var}(y_j^{(t)}) \) for all $i$ and $t$. Since \( y_j^{(t)} \) is a Laplace variable with scale \( C_t \) and each \( \mathbf{v}_i(\tilde{W}) \) is a unit vector, it follows that \( \mathrm{Var}(e_j^{(t)}) = 2 \cdot C_t^2 \).

Using the Chernoff bound and the moment generating function of the Laplacian distribution, we obtain that
\begin{align*}
    \Pr[e_i^{(t)} \geq \sqrt{2} C_t \log n] \leq e^{-\log n} \cdot \mathbb{E}\left[\exp\left({\frac{\sum_j v_{i,j} y_j^{(t)}}{\sqrt{2} C_t}}\right)\right] &= \frac{1}{n} \prod_j \mathbb E\left[\exp\left(\frac{v_{i,j}y_j^{(t)}}{\sqrt2C_t}\right)\right]\\
    &=\frac{1}{n} \prod_j \mathbb E\left[\exp\left(\frac{v_{i,j}}{\sqrt2}\cdot\mathrm{Lap}\left(0,1\right)\right)\right]\\
    &=\frac 1n\prod_j\frac{1}{1 - \frac12v_{i,j}^2}\\
    & \leq \frac{1}{n} \exp{\sum_j v_{i,j}^2} = \frac{e}{n}.
\end{align*}
\end{proof}

Let \( h \) be a positive integer. We discuss the property of the vector $\tilde{W}^h \mathbf{y}^{(t)} := [\gamma_1^{(h,t)}, \dots, \gamma_n^{(h,t)}]^\intercal$ in the next lemma. 
\begin{lemma}
For all \( i,h,t \), the probability that \( |\gamma_i^{(h,t)}| \geq 3\sqrt{2} \cdot \lambda_1(\tilde{W})^h \cdot C_t \cdot \log n  \) is at most \( 2e/n^3 \). \label{lem:sizeofnoise}
\end{lemma}
\begin{proof}
From the definition of \( \gamma_i^{(h,t)} \) and the argument in Lemma~\ref{lem:e}, we find that \( \gamma_i^{(h,t)} = \sum_j \lambda_j(\tilde{W})^h \cdot v_{j,i} \cdot e_j^{(t)} \). According to Lemma \ref{lem:eprop}, $\mathrm{Cov}(e_j^{(t)}, e_{j'}^{(t)}) = 0$ for $j \neq j'$. Therefore, by Lemma \ref{lem:eprop2},
\begin{eqnarray*}
    \mathrm{Var}(\gamma_i^{(h,t)}) & = &  \sum_j \lambda_j(\tilde{W})^{2h} \cdot v_{j,i}^2 \cdot \mathrm{Var}(e_j^{(t)}) \\ & = & 2 C_t^2 \cdot \sum_j \lambda_j(\tilde{W})^{2h} \cdot v_{j,i}^2 \\ & \leq & 2 C_t^2 \cdot \lambda_1(\tilde{W})^{2h} \cdot \sum_j v_{j,i}^2 \\ & = & 2 C_t^2 \cdot \lambda_1(\tilde{W})^{2h}.
\end{eqnarray*}

Since $e_j^{(t)}$ is a linear combination of Laplace variables, $\gamma_i^{(h,t)}$ is also a linear combination of the Laplace variable $y_j^{(t)}$. Let $\mathsf{a}_1, \dots, \mathsf{a}_n$ be real numbers such that $\gamma_i^{(h,t)} = \sum_j \mathsf{a}_j y_j^{(t)}$. We obtain that $\mathrm{Var}(\gamma_i^{(h,t)}) = 2\cdot C_t^2 \sum_j \mathsf{a}_j^2 \leq 2C_t^2 \cdot \lambda_1(\tilde{W})^{2h}$, and $\sum_j \mathsf{a}_j^2 \leq \lambda_1(\tilde{W})^{2h}$. Using the Chernoff bound, we obtain that 
\begin{eqnarray*}
     \Pr[\gamma_i^{(h,t)} \geq 3\sqrt{2} \cdot \lambda_1(\tilde{W})^h \cdot C_t \cdot \log n] 
    & \leq & e^{-3\log n} \cdot \mathbb{E}\left[\exp\left(\frac{\gamma_i^{(h,t)} }{\sqrt{2} \cdot \lambda_1(\tilde{W})^h \cdot C_t} \right)\right] \\
    & \leq & \frac{1}{n^3} \mathbb{E}\left[\exp\left(\frac{\sum_j \mathsf{a}_j \cdot \mathrm{Lap}(0,1)}{\sqrt{2} \cdot \lambda_1(\tilde{W})^h } \right)\right] \\
    & = & \frac{1}{n^3} \prod_j \frac{1}{1 - \frac{\mathsf{a}_j^2}{2\lambda_1(\tilde{W})^{2h}}} \\
    & \leq & \frac{1}{n^3}\exp\left(\frac{1}{2\lambda_1(\tilde{W})^{2h}}\sum_j{\mathsf{a}_j^2}\right) \leq \frac{1}{n^3} \exp(1).
\end{eqnarray*}
The lemma statement follows from the fact that the probability distribution of $\gamma_{h,t}$ is symmetric about $0$.
\end{proof}

In the next lemma, we analyze the size of the  noise added in the algorithm. Recall that the variable $\delta$ is the noisy minimum degree published at Line 2 of Algorithm~\ref{alg:private_modified_PIC}. In Proposition \ref{prop:delta}, we show that $\delta \geq \sqrt{n} \log^4 n$ with probability at least $1 - \frac{1}{n}$. We denote the event that $\delta \geq \sqrt{n} \log^4 n$ by $\mathcal{E}_\delta$.
\begin{lemma}
    Recall that $C_t$ is the scale of the noise added at Line 6 of Algorithm \ref{alg:private_modified_PIC}. Under Assumptions (2) and (4) in Section~\ref{subsec:assumption}, which state that the spectral gap $g$ is constant and the number of nodes $n$ exceeds any fixed constant $C$, we have
    \begin{eqnarray*}\Pr \left[ C_t \leq \frac{10}{9\epsilon} \cdot \frac{\lambda_1(\tilde{W})^{t - 1}}{\sqrt{n} \log^2n} \text{ for all $1 \leq t \leq T$ } \mid \mathcal{E}_\delta\right]  \geq 1 - \frac{8eT^2}{n^2}.\end{eqnarray*} \label{lem:Ct}
\end{lemma}
\begin{proof}
Since \( x_i^{(0)} \) is drawn from a Gaussian distribution with mean \( 0 \) and standard deviation \( 1 \), it follows from the properties of a normal random variable that \( \Pr[|x_i^{(0)}| \geq \log n \cdot \log g] \leq \frac{1}{n^3} \). By applying the union bound, we then have \( \Pr[\max_i |x_i^{(0)}| \geq \log n \cdot \log g] \leq \frac{1}{n^2} \).

We will prove this lemma by induction on the number of iterations \( t \). For \( t = 1 \), recall from Line~6 of the algorithm that the noise \( y_i^{(t)} \) is drawn from a Laplace distribution with scale parameter \( \frac{5 \cdot T}{9 \cdot \epsilon} \cdot \frac{\max_i |x_i^{(t - 1)}|}{\delta} \), where \( \epsilon \) is the privacy budget and \( \delta \) is the minimum degree of the input graph. In the event $\mathcal{E}_\delta$, the variable \( \delta \geq \sqrt{n} \log^4 n \). Recall that we set \( T = 2\frac{\log n}{\log g} \) in our algorithm. 
Consequently, the noise scale in the first iteration is larger than \( \frac{10}{9\epsilon} \frac{\log n}{\log g} \cdot \frac{\log n \cdot \log g}{\sqrt{n} \log^4 n}  = \frac{10}{9 \epsilon \cdot \sqrt{n} \log^2 n} \) with probability not larger than \( 1/n^2 \) when $n$ is large enough.

Next, assume that, in the event $\mathcal{E}_\delta$, with probability not smaller than \( 1 - \frac{2e \cdot (2t - 2)^2}{n^2}  \), for all \( t' < t \), the noise (denoted by \( y_i^{(t')} \)) is sampled from a Laplace distribution with a scale no more than \( \frac{10}{9\epsilon} \cdot \frac{\lambda_1(\tilde{W})^{t' - 1}}{\sqrt{n} \log^2 n} \). From our previous calculations, it follows that \( \mathbf{x}^{(t)} = \tilde{W}^{t} \mathbf{x}^{(0)} + \tilde{W}^{t - 1} \mathbf{y}^{(1)} + \dots + \mathbf{y}^{(t)} \). Let $\tilde{W}^{t} \mathbf{x}^{(0)} = [\mathsf{x}_1^{(t)}, \dots, \mathsf{x}_n^{(t)}]^\intercal$ and, for all $t' \leq t$, $\tilde{W}^{t - t'} \mathbf{y}^{(t')} = [\mathsf{y}_1^{(t,t')}, \dots, \mathsf{y}_n^{(t,t')}]^\intercal$. The value of $\max_i |x_i^{(t - 1)}|$, which decides the noise scale of $\mathbf{y}^{(t)}$, is equal to $\max_i \left|\mathsf{x}_i^{(t - 1)} + \sum\limits_{t' = 1}^{t - 1} \mathsf{y}_i^{(t - 1,t')}\right|$.

Let us first consider the vector \( [\mathsf{x}^{(t - 1)}_1, \dots, \mathsf{x}^{(t - 1)}_n]^\intercal \). Recall that \( \mathbf{v}_i(\tilde{W}) = [v_{i,1}, \dots, v_{i,n}]^\intercal \). By the notation, we have \( \mathsf{x}_{i}^{(t - 1)} = \sum_j \lambda_j(\tilde{W})^{t - 1} v_{j,i} c_j \).

Since, by Lemma \ref{lem:c}, $c_j$ and $c_{j'}$ are independent for $j \neq j'$, we obtain:
\begin{eqnarray*}
\mathbb{E}[\mathsf{x}_{i}^{(t - 1)}] & = & \sum_j \lambda_j(\tilde{W})^{t - 1} v_{j,i} \cdot \mathbb{E}[c_j] = 0, \\
\mathrm{Var}[\mathsf{x}_{i}^{(t - 1)}] & = & \sum_j \lambda_j(\tilde{W})^{2t - 2} v_{j,i}^2 \mathrm{Var}[c_j] \leq \lambda_1(\tilde{W})^{2t - 2} \mathrm{Var}\left[\sum_j v_{j,i} c_j\right]  \\ & = &  \lambda_1(\tilde{W})^{2t - 2} \mathrm{Var}\left[x_i^{(0)}\right] = \lambda_1(\tilde{W})^{2t - 2}.
\end{eqnarray*}
Also, since $\mathsf{x}_i^{(t - 1)}$ is a linear combination of the normal random variable $c_j$, we can conclude that $\mathsf{x}_i^{(t - 1)}$ is also normal. By the property of the normal variable, we have $\Pr\left[|\mathsf{x}_i^{(t - 1)}| \geq \frac{1}{2} \log n \cdot \log g \cdot \lambda_1(\tilde{W})^{t - 1}\right] \leq \frac{1}{n^3}$ for all $i$. By the union bound, $\Pr\left[\max_i |\mathsf{x}_i^{(t - 1)}| \geq \frac{1}{2} \log n \cdot \log g \cdot \lambda_1(\tilde{W})^{t - 1}\right] \leq \frac{1}{n^2}$. 

Let us reconsider the variable \(\gamma_i^{(h,t)}\) from Lemma \ref{lem:sizeofnoise}. Note that \(\mathsf{y}_i^{(t - 1, t')} = \gamma^{(t - t' - 1, t')}\). Let $\mathcal{E}$ denote the event that \(C_{t'} \leq \frac{10}{9\epsilon} \cdot \frac{\lambda_1^{t' - 1}(\tilde{W})}{\sqrt{n} \log^2 n}\) for all \(t' < t\).
In the event $\mathcal{E}$ and $\mathcal{E}_\delta$, Lemma \ref{lem:sizeofnoise} implies that, for all \(i, t, t'\),  
\begin{eqnarray*}
\left| \mathsf{y}_i^{(t - 1, t')} \right|& \geq & 3\sqrt{2} \cdot \lambda_1^{t - t' - 1}(\tilde{W}) \cdot \frac{10}{9\epsilon} \cdot \frac{\lambda_1^{t' - 1}(\tilde{W})}{\sqrt{n} \log^2 n}  = \frac{30\sqrt{2}}{9\epsilon} \cdot \frac{\lambda_1^{t - 2}(\tilde{W})}{\sqrt{n} \log^2 n} 
\end{eqnarray*}
with probability at most \(\frac{2e}{n^3}\).

By applying the union bound, we deduce that for all \(t, t'\),  
\[
\max_i |\mathsf{y}_i^{(t - 1, t')}| \geq \frac{30\sqrt{2}}{9\epsilon} \cdot \frac{\lambda_1^{t - 2}(\tilde{W})}{\sqrt{n} \log^2 n}
\]  
with probability at most \(\frac{2e}{n^2}\). By Lemma 4.4 of \citep{mohar1989isoperimetric}, we have that $\lambda_2(B) \geq 0$ and $\lambda_1(\tilde{W}) \geq \frac{1}{2}$. 
When \(n\) is sufficiently large, it follows that, for all \(t, t'\),  
\[
\max_i |\mathsf{y}_i^{(t - 1, t')}| \geq \frac{  \lambda_1^{t - 1}(\tilde{W}) \log g}{4 \log n} \geq \frac{30 \sqrt{2}}{9\epsilon}\frac{\lambda_1^{t - 2}(\tilde{W})}{\sqrt{n} \log^2 n}
\]  
with probability at most \(\frac{2e}{n^2}\). We finally obtain  
$$\Pr\left[\sum_{t' \leq t - 1} \max_i |\mathsf{y}_i^{(t - 1, t')}| \geq \frac{1}{2} \cdot \lambda_1^{t - 1}(\tilde{W}) \mid \mathcal{E}, \mathcal{E}_\delta \right] \leq \frac{2et}{n^2}.$$  

Because, for all $i$ and $t$, the variables $\mathsf{x}_i^{(t - 1)}$ do not depends on the scale of the Laplacian noise and the event $\mathcal{E}$, we obtain that:
\begin{eqnarray*}
& & \Pr\left[\max_i \left|x_i^{(t-1)}\right| \geq \log n \cdot \log g \cdot \lambda_1(\tilde{W})^{t - 1} \mid \mathcal{E}, \mathcal{E}_\delta \right] \\
    & = & \Pr\left[\max_i \left|\mathsf{x}_i^{(t - 1)} + \sum_{t' \leq t - 1} \mathsf{y}_i^{(t - 1,t')}\right| \geq \log n \cdot \log g \cdot \lambda_1(\tilde{W})^{t - 1} \mid \mathcal{E}, \mathcal{E}_\delta\right] \\
    & \leq & \Pr\left[\max_i \left|\mathsf{x}_i^{(t - 1)}\right| + \sum_{t' \leq t - 1} \max_i \left| \mathsf{y}_i^{(t - 1,t')}\right| \geq \log n \cdot \log g\cdot \lambda_1(\tilde{W})^{t - 1} \mid \mathcal{E}, \mathcal{E}_\delta\right] \\
    & \leq & \Pr \left[\max_i \left|\mathsf{x}_i^{(t - 1)}\right| \geq \frac{1}{2} \log n \cdot \log g \cdot \lambda_1(\tilde{W})^{t - 1} \right] \\ & & ~~~~ + \Pr \left[\sum_{t' \leq t - 1} \max_i \left|\mathsf{y}_i^{(t - 1,t')}\right| \geq \frac{1}{2}\lambda_1(\tilde{W})^{t - 1} \mid \mathcal{E}, \mathcal{E}_\delta \right] \\ & \leq & \frac{(2et + 1)}{n^2}.
\end{eqnarray*}
In the event $\mathcal{E}$ and $\mathcal{E}_\delta$, $\max_i |x_i^{(t-1)}| \geq \log n \cdot \log g \cdot \lambda_1(\tilde{W})^{t - 1}$ with probability at most $\frac{2et + 1}{n^2}$. In the event of $\mathcal{E}$ and $\mathcal{E}_\delta$, the noise scale at the iteration $t$, denoted by $C_t$, is at most $\frac{10}{9\epsilon} \frac{2\log n}{\log g}  \frac{\log n \cdot \log g \cdot \lambda_1(\tilde{W})^{t - 1}}{\sqrt{n} \log^4 n} = \frac{10}{9\epsilon} \cdot \frac{\lambda_1(\tilde{W})^{t - 1}}{\sqrt{n} \log^2 n}$ with probability at least $1 - \frac{2et + 1}{n^2}$. As a result,
\begin{eqnarray*}
 \Pr\left[C_t \geq \frac{10}{9\epsilon} \cdot \frac{\lambda_1(\tilde{W})^{t - 1}}{\sqrt{n} \log^2 n} \text{ or } \bar{\mathcal{E}} \mid \mathcal{E}_\delta \right] 
& \leq & \Pr\left[C_t \geq \frac{10}{9\epsilon} \cdot \frac{\lambda_1(\tilde{W})^{t - 1}}{\sqrt{n} \log^2 n} \text{ and } \mathcal{E} \mid \mathcal{E}_\delta \right] + \Pr[\bar{\mathcal{E}} \mid \mathcal{E}_\delta] \\
& \leq & \Pr\left[C_t \geq \frac{10}{9\epsilon} \cdot \frac{\lambda_1(\tilde{W})^{t - 1}}{\sqrt{n} \log^2 n} \mid \mathcal{E}, \mathcal{E}_\delta \right] +  \frac{2e(2t - 2)^2}{n^2} \\
& \leq & \frac{2et + 1}{n^2} + \frac{2e(2t - 2)^2}{n^2} \leq \frac{2e(2t)^2}{n^2}.
\end{eqnarray*}
This completes the induction step. We can conclude that, for all $t \in \{1, \dots, T \}$, 
$C_{t'} \leq \frac{10}{9\epsilon} \cdot \frac{\lambda_1(\tilde{W})^{t' - 1}}{\sqrt{n} \log^2 n}$ for all $t' \leq t$ with probability at least
$1 - \frac{2e(2t)^2}{n^2}$ when $\delta \geq \sqrt{n} \log^4 n$.
\end{proof}

We will leverage the previous lemma to demonstrate that the outcome of Algorithm \ref{alg:private_modified_PIC} closely aligns with the results obtained through spectral clustering. Recall Lemma \ref{lem:e} that the final vector $x_i^{(T)} = \sum_{j = 1}^n \tilde{c}_j v_{j,i}$ when $\tilde{c}_j = c_j \lambda_j(\tilde{W})^T + \sum_{t = 1}^T e_j^{(t)} \lambda_j(\tilde{W})^{T - t}$.
\begin{theorem}
   Under Assumptions (1)–(4) in Section~\ref{subsec:assumption}, which state that (1) the minimum degree is at least $2\sqrt{n} \log^4 n$, (2) the spectral gap $g$ is constant, (3) there exist constants $\delta \approx 1$ and $\gamma < 1$ such that the components of $\mathbf{v}_2(B)$ satisfy $\left|\left\{i : |\nu_i| > \frac{\gamma}{\sqrt{n}} \right\}\right| \geq \delta \cdot n$, and (4) the number of nodes $n$ exceeds any fixed constant $C$, it follows that for any node $i$ with $|v_{1,i}| \geq \frac{\gamma}{\sqrt{n}}$, and sufficiently large $n$, we have
        $$\Pr\left[\left|c_1 \lambda_1(\tilde{W})^T v_{1,i}\right| > \left|\sum_{t =1}^Te_1^T\lambda_1(\tilde{W})^{T - t}v_{1,i} + \sum_{j = 2}^n \tilde{c}_j v_{j,i} \right| \right] \geq 0.95 - o(1).$$
    \label{thm:precision}
\end{theorem}
\begin{proof}
    We first obtain that
\begin{eqnarray}
    & & \Pr\left[\left|c_1 \lambda_1(\tilde{W})^T v_{1,i}\right| > \left|\sum_{t =1}^Te_1^T\lambda_1(\tilde{W})^{T - t}v_{1,i} + \sum_{j = 2}^n \tilde{c}_j v_{j,i} \right| \right] \nonumber \\ & \geq & \Pr\left[\left|c_1 \lambda_1(\tilde{W})^T v_{1,i}\right| > \left|\sum_{t =1}^Te_1^T\lambda_1(\tilde{W})^{T - t}v_{1,i}\right| + \left|\sum_{j = 2}^n \tilde{c}_j v_{j,i} \right| \right] \nonumber \\
   & \geq & \Pr \Bigg[ |v_{1,i}| \left( |c_1  \lambda_1(\tilde{W})^T| - \left| \sum_{t = 1}^T e_1^{(t)} \lambda_1(\tilde{W})^{T - t}  \right| \right) \nonumber \\ & & ~~~~~~~~~ > \left|\sum_{j = 2}^n c_j v_{j,i} \lambda_j(\tilde{W})^T \right|  + \left| \sum_{j = 2}^{n} \sum_{t = 1}^T e_j^{(t)} \lambda_j(\tilde{W})^{T - t}v_{j,i}\right| \Bigg] \nonumber. \label{eqn:33} \end{eqnarray}
 Recall from Lemma \ref{lem:c} that $c_i$ is a normal random variable with mean 0 and standard deviation 1. We obtain that:
 \begin{equation}
 \Pr\left[ \left|c_1 \lambda_1(\tilde{W})^T\right| \geq \frac{\lambda_1(\tilde{W})^T}{16} \right] > 0.95. \label{eqn:0.95}
 \end{equation}

 Recall from Lemma \ref{lem:eprop2} that $\Pr\left[e_1^{(t)} \geq \sqrt{2} C_t \log n\right] \leq \frac{e}{n}$.
 Let $\mathcal{E}$ be the event that $\max_t C_t \leq \frac{10}{9\epsilon}\cdot \frac{\lambda_1(\tilde{W})^{t - 1}}{\sqrt{n} \log^2 n}$ and $\mathcal{E}_\delta$ be the event that $\delta \geq \sqrt{n} \log^4 n$. We obtain that $\Pr\left[e_1^{(t)} \geq \frac{10\sqrt{2}}{9\epsilon} \frac{\lambda_1(\tilde{W})^{t - 1}}{\sqrt{n} \log n} 
 | \mathcal{E}, \mathcal{E}_\delta \right] \leq \frac{e}{n}$. Denote $\sum_{t = 1}^{T} e_j^{(t)} \lambda_1(\tilde{W})^{T - t}$ by $\eta$. By the union bound,
 \begin{eqnarray*}
 \Pr\left[\eta \geq \frac{\lambda_1(\tilde{W})^T}{32} | \mathcal{E} \right]
 \leq \Pr\left[\eta \geq \frac{10 \sqrt{2} \cdot T}{9\epsilon} \frac{\lambda_1(\tilde{W})^{T - 1}}{\sqrt{n} \log n} | \mathcal{E} \right]  \leq \frac{e T}{n},
 \end{eqnarray*}
 for sufficiently large $n$. Recall the event $\mathcal{E}_\delta$, which is the event when $\delta \geq \sqrt{n} \log^4 n$. By Lemma \ref{lem:Ct}, we know that $\Pr[\mathcal{E} \mid \mathcal{E}_\delta] \geq 1 - \frac{8eT^2}{n^2}$. Also, by Proposition \ref{prop:delta}, we know that $\Pr[\bar{\mathcal{E}_\delta}] \leq \frac{1}{2n}$. As a result, when $n$ is large enough,
 \begin{eqnarray*}
\Pr\left[\eta \geq \frac{\lambda_1(\tilde{W})^T}{32} \right] \nonumber
& \leq & \Pr\left[\eta \geq \frac{\lambda_1(\tilde{W})^T}{32} \text{ and } \mathcal{E} \text{ and } \mathcal{E}_\delta \right]  + \Pr[\bar{\mathcal{E}} \text{ and } \mathcal{E}_\delta] + \Pr[\bar{\mathcal{E}_\delta}] \nonumber \\
& \leq & \Pr\left[ \eta \geq \frac{\lambda_1(\tilde{W})^T}{32} \mid \mathcal{E}, \mathcal{E}_\delta \right] +  \Pr[\bar{\mathcal{E}} \mid \mathcal{E}_\delta] + \frac{1}{2n} \nonumber = o(1). 
\end{eqnarray*}
Since the distribution of $\eta$ is symmetric around 0, we have that
\begin{equation}
    \Pr\left[|\eta| \geq \frac{\lambda_1(\tilde{W})^T}{32}  \right] = o(1).
    \label{eqn:eta}
\end{equation}
By combining (\ref{eqn:0.95}) and (\ref{eqn:eta}) and using the fact that $|v_{1,i}| \geq \frac{\gamma}{\sqrt{n}}$, we obtain that:
\begin{equation}
\Pr\left[ |v_{1,i}| \left( |c_1  \lambda_1(\tilde{W})^T| - \left| \eta  \right| \right) \geq \frac{\gamma \lambda_1(\tilde{W})^T}{32 \sqrt{n}}\right] > 0.95 - o(1) \label{eqn:left-side}
\end{equation}

By the assumption that $\lambda_2(\tilde{W}) \leq \frac{\lambda_1(\tilde{W})}{g}$, we have that $\left| \sum_{j = 2}^n c_j v_{j,i} \lambda_j(\tilde{W})^T  \right| \leq \left| \frac{\lambda_1(\tilde{W})^T}{g^T} \cdot \sum_{j = 2}^n c_j \right|$. Since $\sum_{j = 2}^n c_j v_{j,i} \lambda_j(\tilde{W})^T$ is a normal random variable with mean 0 and standard deviation at most $\frac{\sqrt{n} \cdot \lambda_1(\tilde{W})^T}{g^T}$, we have the following.
 \begin{equation*}
 \Pr\left[ \left|\sum_{j = 2}^n c_j v_{j,i} \lambda_j(\tilde{W})^T\right| \geq \frac{\sqrt{n} \log n \cdot \lambda_1(\tilde{W})^T}{g^T} \right] < \frac{1}{n^2}.
 \end{equation*}
 When $T = \frac{2 \log n}{\log g}$ and $n$ is large enough, we obtain that $\frac{\sqrt{n} \log n}{g^T} < \frac{\gamma}{65 \sqrt{n}}$. Hence,
 \begin{equation}
 \Pr\left[ \left|\sum_{j = 2}^n c_j v_{j,i} \lambda_j(\tilde{W})^T\right| \geq \frac{\gamma \cdot \lambda_1(\tilde{W})^T}{65 \sqrt{n}} \right] < \frac{1}{n^2}. \label{eqn:right-side-1}
 \end{equation}

 Consider the summation $\sum_{j = 2}^n \sum_{t = 1}^T e_j^{(t)} \lambda_j(\tilde{W})^{T - t} v_{j,i}$. Denote the summation as $\beta_t$. By Lemma~\ref{lem:eprop}, we know that $\mathbb{E}[e_j^{(t)}] = 0$ for all $j, t$. As a result, $\mathbb{E}[\beta_t] = \mathbb{E}\left[\sum_{j = 2}^n e_j^{(t)} \lambda_j(\tilde{W})^{T - t} v_{j,i}\right] = 0$ for all $t$. Furthermore, by $\lambda_j(\tilde{W}) \leq \frac{\lambda_1(\tilde{W})}{g}$ for all $j \geq 2$,
 \begin{eqnarray}
      \mathrm{Var}\left[\beta_t\right] \nonumber & \leq & \sum_{j = 2}^n \lambda_j(\tilde{W})^{2T - 2t} v_{j,i}^2 \mathrm{Var}(e_j^{(t)})\nonumber \\ & \leq & \sum_{j = 1}^n  \frac{\lambda_1(\tilde{W})^{2T - 2t}}{g^{2T - 2t}} v_{j,i}^2 \mathrm{Var}(e_j^{(t)}) \nonumber \\
     & = & \frac{\lambda_1(\tilde{W})^{2T - 2t}}{g^{2T - 2t}} \mathrm{Var}\left[\sum_j v_{j,i} e_j^{(t)}\right] \nonumber \\
     & = & \frac{\lambda_1(\tilde{W})^{2T - 2t}}{g^{2T - 2t}} \mathrm{Var}\left[y_i^{(t)}\right] \nonumber \\ & = & 2 \frac{\lambda_1(\tilde{W})^{2T - 2t}}{g^{2T - 2t}} C_t^2. \label{eqn:5}
 \end{eqnarray}
We observe that both \(e_j^{(t)}\) and \(\beta_t = \sum_{j=2}^n e_j^{(t)} \lambda_j(\tilde{W})^{T-t} v_{j,i}\) can be written as linear combinations of \(y_1^{(t)}, \dots, y_n^{(t)}\). Let \(\mathsf{b}^{(t)}_1, \dots, \mathsf{b}^{(t)}_n\in\mathbb{R}\) be such that
$\beta_t
\,=\,
\sum_{j=1}^n \mathsf{b}^{(t)}_j\,y_j^{(t)}$. By~(\ref{eqn:5}), $\mathrm{Var}\left[\beta_t \right]
\,=\,
2 \sum_{j=1}^n \mathsf{b}_j^2 C_t^2 \leq 2 \frac{\lambda_1(\tilde{W})^{2T - 2t}}{g^{2T - 2t}} C_t^2$ and $\sum_{j=1}^n \mathsf{b}_j^2 \leq \frac{\lambda_1(\tilde{W})^{2T - 2t}}{g^{2T - 2t}}$. Using the Chernoff bound and the moment generating function of the Laplacian distribution, we obtain that
\begin{eqnarray*}
    \Pr\left[\beta_t \geq \sqrt{2} \log n \cdot \frac{\lambda_1^{T-t}(\tilde{W})}{g^{T - t}} \cdot C_t\right] 
    & \leq & e^{-\log n} \cdot \mathbb{E}\left[\exp \left(\frac{\beta_t}{\sqrt{2} \cdot \frac{\lambda_1^{T-t}(\tilde{W})}{g^{T - t}} \cdot C_t} \right) \right] \\
    & \leq & \frac{1}{n} \mathbb{E}\left[\exp \left( \frac{\sum_j \mathsf{b}_j \mathrm{Lap}(0,1)}{\sqrt{2} \cdot \frac{\lambda_1^{T-t}(\tilde{W})}{g^{T - t}}}\right)\right] \\
    & = & \frac{1}{n} \prod_j \frac{1}{1 -  \mathsf{b}_j^2 / \left(2 \cdot \frac{\lambda_1^{2T-2t}(\tilde{W})}{g^{2T - 2t}}\right) } \\
    & \leq & \frac{1}{n}\exp\left(\frac{g^{2T - 2t}}{\lambda_1^{2T-2t}(\tilde{W})}\sum_j{\mathsf{b}_j^2}\right) \leq \frac{e}{n}.
\end{eqnarray*}
 
 Let $\mathcal{E}$ be the event that $\max_t C_t \leq \frac{10}{9\epsilon} \cdot \frac{\lambda_1(\tilde{W})^{t - 1}}{\sqrt{n} \log^2 n}$. For large $n$ such that $\log n \geq \frac{650 \sqrt{2}}{9\epsilon} \cdot \frac{g}{g - 1} \cdot \frac{1}{\gamma}$,
 \begin{eqnarray*}
     \Pr\left[ \beta_t \geq \frac{\gamma  \lambda_1(\tilde{W})^T}{65 \cdot \frac{g}{g-1} \sqrt{n} g^{T - t}} \mid \mathcal{E} \right]  \leq \Pr\left[\beta_t \geq \frac{10\sqrt{2}}{9\epsilon} \cdot \frac{ \lambda_1^{T-1}(\tilde{W})}{\log n \sqrt{n} g^{T - t}} \mid \mathcal{E} \right]  \leq \frac{e}{n}.
 \end{eqnarray*}
 Recall that $\mathcal{E}_\delta$ is the event such that $\delta \geq \sqrt{n} \log^4 n$. By Lemma \ref{lem:Ct}, we know that $\Pr[\mathcal{E} \mid \mathcal{E}_\delta] = 1 - \frac{8eT^2}{n^2}$, and, by Proposition \ref{prop:delta}, we know that $\Pr[\bar{\mathcal{E}_\delta}] \leq \frac{1}{2n}$.
 As a result, when $n$ is large enough,
\begin{eqnarray*}
\Pr\left[\beta_t \geq \frac{\gamma  \lambda_1(\tilde{W})^T}{65 \cdot \frac{g}{g-1}\sqrt{n} g^{T - t}}\right] \nonumber 
& \leq & \Pr\left[\beta_t \geq \frac{\gamma  \lambda_1(\tilde{W})^T}{65 \cdot \frac{g}{g-1}\sqrt{n}g^{T - t}} \text{ and } \mathcal{E} \text{ and } \mathcal{E}_\delta \right] \\ & & ~~~~ + \Pr[\bar{\mathcal{E}} \text{ and } \mathcal{E}_\delta] + \Pr[\mathcal{E}_\delta] \nonumber \\
& \leq & \Pr\left[ \beta_t \geq \frac{\gamma  \lambda_1(\tilde{W})^T}{65 \cdot \frac{g}{g-1}\sqrt{n}g^{T - t}} \mid \mathcal{E} \right] + \Pr\left[\bar{\mathcal{E}} \mid \mathcal{E}_\delta \right] + \frac{1}{2n} \nonumber \\ & = & O\left(\frac{1}{n}\right). 
\end{eqnarray*}
 By the union bound and by $\sum_{t = 1}^T \frac{1}{g^{T - t}} = \frac{1}{g^T} \sum_{t = 1}^{T} g^{t} = \frac{1}{g^T} \frac{g^{T + 1} - 1}{g - 1} = \frac{gn^2 - 1}{gn^2 - n^2} \leq \frac{g}{g - 1}$, we obtain that 
 \begin{eqnarray*}
     \Pr\left[\sum_{t = 1}^{T} \beta_t \geq \frac{\gamma  \lambda_1(\tilde{W})^T}{65\sqrt{n}} \right] 
      \leq \Pr\left[\sum_{t = 1}^{T} \beta_t \geq \sum_{t = 1}^T \frac{\gamma  \lambda_1(\tilde{W})^T}{65 \cdot \frac{g}{g-1} \cdot \sqrt{n} g^{T - t}} \right] = O\left(\frac{\log n}{n}\right). 
 \end{eqnarray*}
 As $\sum_{t = 1}^T \beta_t$ is a linear combination of Laplacian random variable, we know that the distribution of the summation is symmetric around $0$. Hence, 
 \begin{equation}
     \Pr\left[\left| \sum_{j = 2}^n \sum_{t = 1}^T e_j^{(t)} \lambda_j(\tilde{W})^{T - t} v_{j,i} \right| \geq \frac{\gamma  \lambda_1(\tilde{W})^T}{65\sqrt{n}} \right] = o(1). \label{eqn:right-side-2}
 \end{equation}

We then obtain the lemma statement by combining (\ref{eqn:left-side}) with inequalities (\ref{eqn:right-side-1}) and (\ref{eqn:right-side-2}).
\end{proof}

\section{Further experiments}
\label{appendix:exp}

In this appendix, we present additional experimental results, specifically demonstrating that each modification introduced in Algorithm \ref{alg:private_modified_PIC} contributes to improved precision.

\subsection{Results of using the lazy random walk matrix in the iterative spectral clustering (Key contribution 2 in Section \ref{sec:alg})}
In this subsection, we analyze the effect of performing power iteration with the lazy random walk matrix $W_\alpha = \alpha I + (1-\alpha)D^{-1}A$ instead of the usual random walk matrix $D^{-1}A$ used during PIC \citep{PIC-LinCohen-2010, PIC-Provably-BoustidisKambadurGittens-2015}.

We start with an $n$-dimensional standard normal variable $\mathbf x^{(0)}$, and iteratively obtain $\mathbf x^{(t)}=W_\alpha\cdot \mathbf x^{(t-1)}-\frac1n\sum_i\mathbf x^{(t-1)}_i$.
This is equivalent to the PIC algorithm with $k=2$ initial vectors.
\begin{figure}[ht]
\centering
\includegraphics[width=0.6\linewidth]{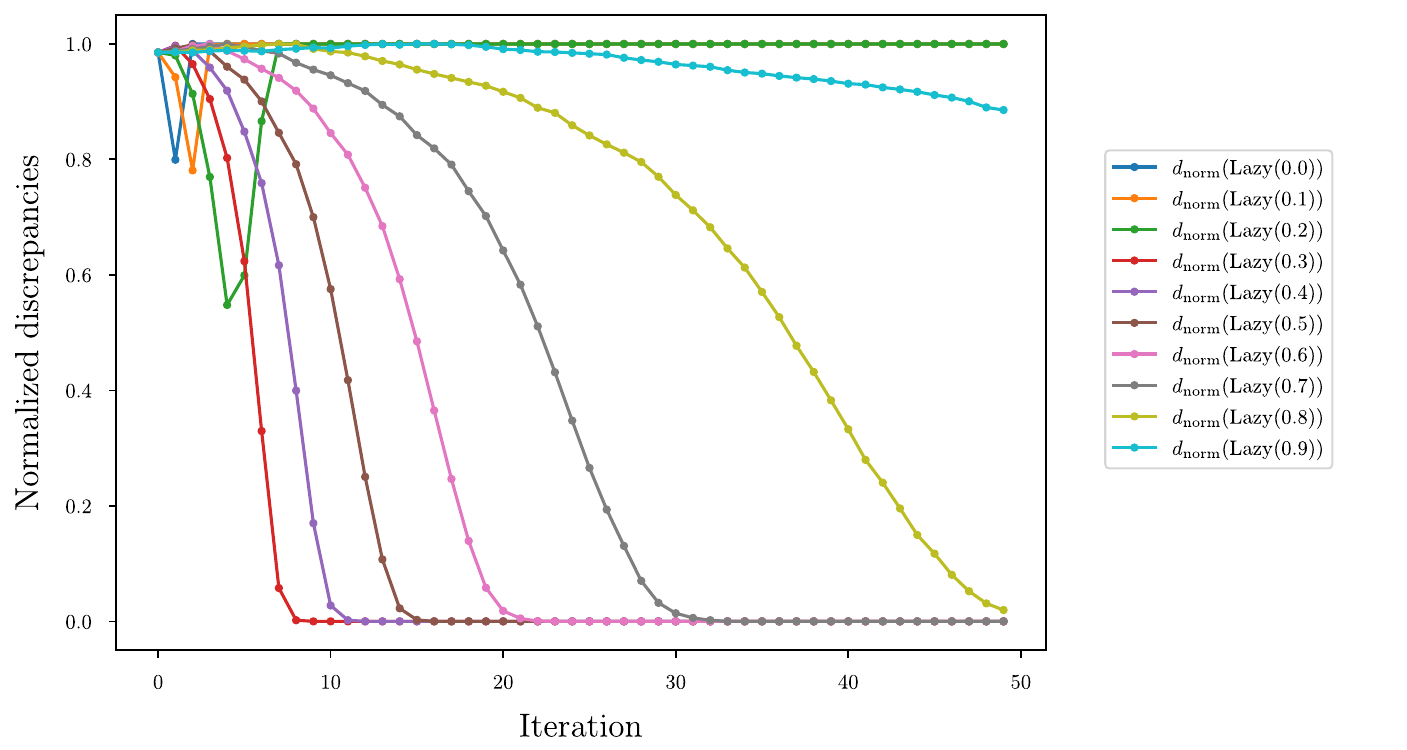}
\caption{Power iteration on $\mathrm{BSBM}(1000, 1000, 1000, 1000, 0.5, 0.2)$ for lazy random walk matrices $W_\alpha$ with $\alpha\in\{0, 0.1,\ldots, 0.9\}$. \label{fig:lazy_experiment_artificial}}
\end{figure}

For bipartite graphs, the random walk matrix $W_0=D^{-1}A$ has $-1$ as an eigenvalue.
Thus, for bipartite $2$-clustered graphs, the performance of PIC is not good unless more initial vectors are selected.
We demonstrate this by introducing a \emph{Bipartite Stochastic Block Model} graph with two clusters, defined as follows.
Given integers $a_i, b_i$ and probabilities $p$ and $q$, a graph $G\sim\mathrm{BSBM}(a_1, a_2, b_1, b_2, p, q)$ has node set $A_1\sqcup A_2\sqcup B_1\sqcup B_2$ with $|A_i|=a_i$ and $|B_i|=b_i$, such that every pair of nodes between $A_i$ and $B_i$ is added with probability $p$, and $A_i$ and $B_j$ are added with probability $q$.
This graph is bipartite with independent sets $A_1\cup A_2$ and $B_1\cup B_2$, and when $p\gg q$, admits a clear cluster structure given by the node clusters $A_1\cup B_1$ and $A_2\cup B_2$.

Figure~\ref{fig:lazy_experiment_artificial} shows that for certain BSBM's, the produced clusters by iteratively multiplying $W_\alpha$ always have discrepancy close to $1$ when $\alpha$ is close to $0$.
On the other hand, when $\alpha\approx1$, the procedure is too slow to converge since $W_\alpha\approx I$.
Therefore, selecting a lazy factor of $\alpha=\frac12$ seems to be a natural choice in general when no additional information about the input graph is available.

\subsection{Result of leading eigenvector elimination (Key contribution 1 in Section \ref{sec:alg})}

\begin{figure}[ht]
\centering
\includegraphics[width=0.8\linewidth]{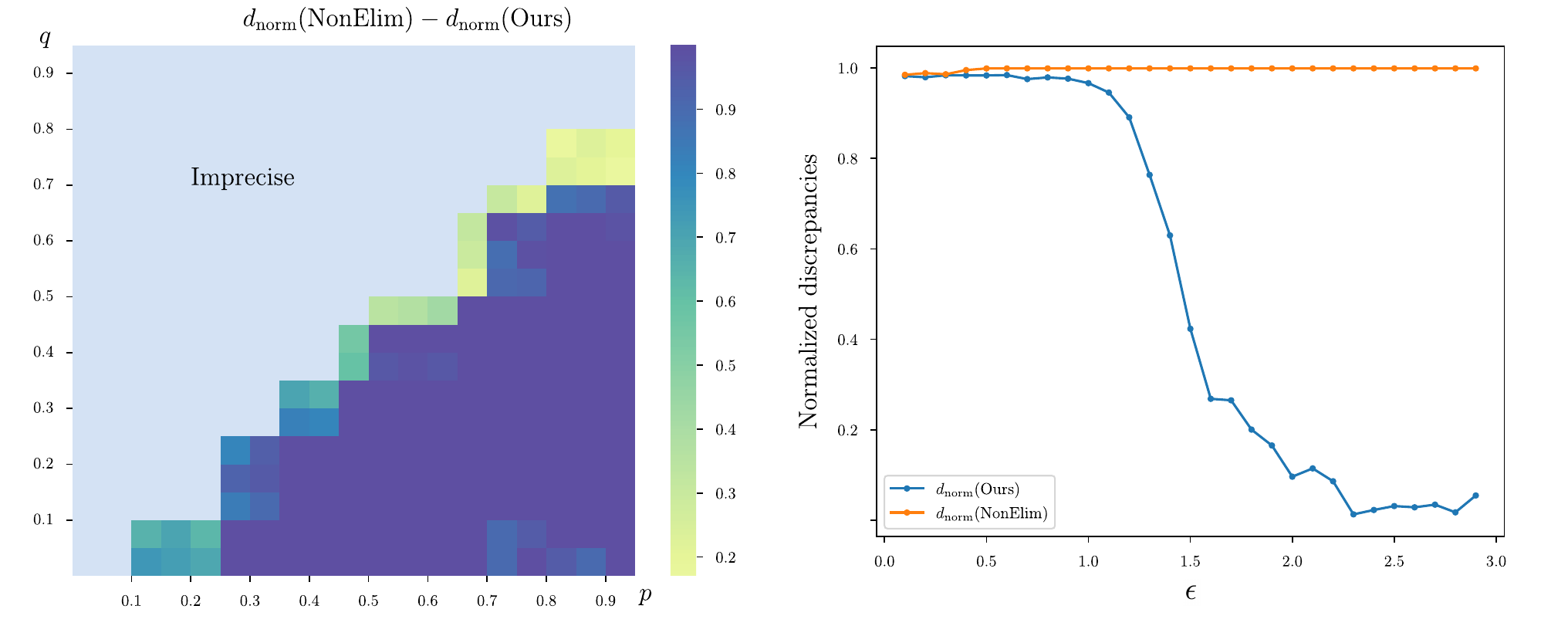}
\caption{(Left): Heatmap of average $d_{\mathrm{norm}}(\text{NonElim})-d_{\mathrm{norm}}(\text{Ours})$ over $20$ SBMs with $n_1=n_2=1000$, with varying probabilities $p,q\in\{0.05,0.1,\ldots, 0.95\}$, and privacy budget $\epsilon=2.0$.
(Right): Discrepancy with increasing $\epsilon$ for 20 SBMs with $p=0.3, q=0.2$. \label{fig:comparison_with_NonElim}}
\end{figure}

Now, we perform an experiment to investigate the effect of elimination of the leading eigenvalue of the lazy random walk matrix, which is a procedure changing the matrix $W$ to the matrix $\tilde{W}$ described in Difference 4 of Section \ref{sec:alg}. For this experiment, we select $\varepsilon=2.0$ and generate $20$ SBM's of cluster sizes $1000$ for pairs of probabilities $(p,q)$.

We present our results in Figure \ref{fig:comparison_with_NonElim}. The figure illustrates that our algorithm, without the leading vector elimination (referred to as NonElim), consistently fails to recover the original clusters. Although not depicted in the graph, we observed that the NonElim algorithm fails to successfully identify the clusters even when the privacy budget $\epsilon$ is set as high as $20$. In contrast, our proposed method successfully identifies these clusters with minimal discrepancy.

\subsection{Results on degree-corrected stochastic block model}
\label{appedix:resultsDCBM}
We also run our algorithm for the degree-corrected stochastic block model (DCBM) \citep{karrer-newman-DCBM-2011}.
This synthetic model provides a correction to the stochastic block model such that the graphs generated from DCBM have an expected degree sequence given by $(\theta_1,\ldots,\theta_n)$.
Let $B_{\text{sbm}(p,q,n_1,n_2)}$ be the expected adjacency matrix of the usual SBM with probability of edges inside the same cluster equal to $p$ and between different clusters equal to $q$, $n_1$ and $n_2$ nodes in the two clusters, respectively.
If $\theta=\text{diag}(\theta_1,\ldots,\theta_n)$, the DCBM has the edge-probability matrix given by $\theta B_{\text{sbm}(p,q,n_1,n_2)} \theta$.
Since it is known that the degree sequences of many social networks follow the power law \citep{bollobas-ScaleFreePowerLaw-2001, choromanski2013scale}, we make sure that our generated graphs follow the power law, i.e. the number of $\theta_i$ with value $d$ is proportional to $d^{-\alpha}$ for some $\alpha\in(2,3)$.

\begin{figure}[t]
    \centering
    \subfigure[Comparison across different privacy budget]{%
        \includegraphics[width=0.47\textwidth]{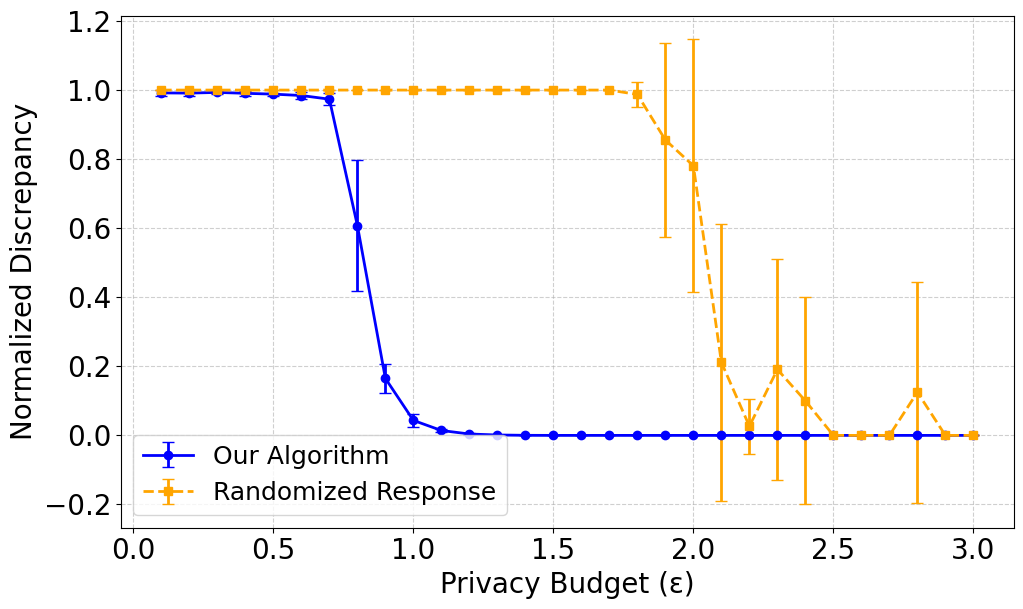} 
        \label{subfig:dcbm-varyeps}
    }
    \hfill
    \subfigure[Comparison across different graph sizes]{%
        \includegraphics[width=0.47\textwidth]{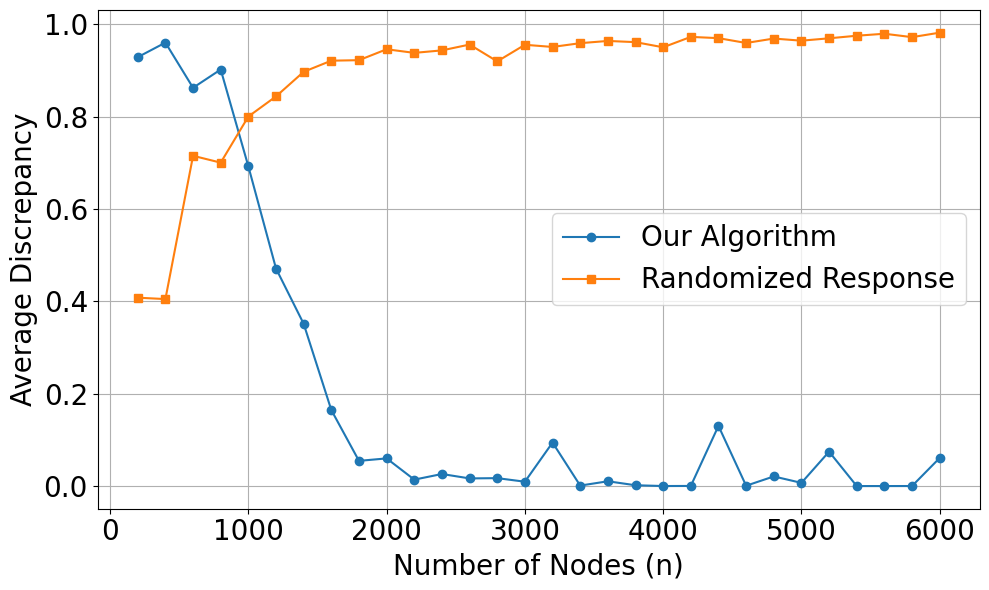} 
        \label{subfig:dcbm-varynumnodes}
    }
    \caption{
The normalized discrepancies of our algorithm for the degree-corrected stochastic block model following a power law in the degree distribution}
    \label{fig:dcbm-results}
\end{figure}

In our experiment, we select $\alpha=2.5$ and generate $10$ different degree sequences $\theta$, with the additional constraint that $\min\theta\ge 5\sqrt{n}$, where $n$ is the number of nodes of the DCBM to be generated.
For each $\theta$, we permute the node labels and generate DCBM graphs with parameters $p=0.4$, $q=0.1$.
Due to memory restrictions, we generate $10$ DCBM graphs per degree sequence for Figure~\ref{subfig:dcbm-varyeps}, and a single DCBM per degree sequence for Figure~\ref{subfig:dcbm-varynumnodes}.
In Figure~\ref{subfig:dcbm-varyeps}, we plot the average normalized discrepancies with the randomized response and our proposed algorithm for varying privacy budgets $\varepsilon$, and in Figure~\ref{subfig:dcbm-varynumnodes} we plot the same for varying number of nodes $n$ (with each cluster having same size).

We observe similar trends in the results obtained from graphs generated using the SBM, as presented in Section~\ref{sec:exp}. Our algorithm consistently outperforms the randomized response mechanism across various privacy budgets. When varying the graph size, randomized response performs better on smaller graphs; however, our method achieves lower discrepancy when the number of nodes is at least 1000.

\subsection{Additional results on Reddit graphs}
\label{appedix:additionalReddit}

We present our experimental results for the 500-core and 100-core decompositions of the Reddit graph in this section. The 500-core decomposition includes 44,586 nodes and 54,984,204 edges, while the 100-core decomposition includes 154,525 nodes and 108,024,958 edges. The experimental settings are identical to those described in Section~\ref{subsec:reddit}.

Due to memory limitations, we were unable to run the randomized response algorithm on these decompositions, even using an A100 GPU with 40GB of GPU memory and 83.5GB of system RAM.  

\begin{figure}[t]
    \centering
    \subfigure[500-core decomposition]{%
        \includegraphics[width=0.47\textwidth]{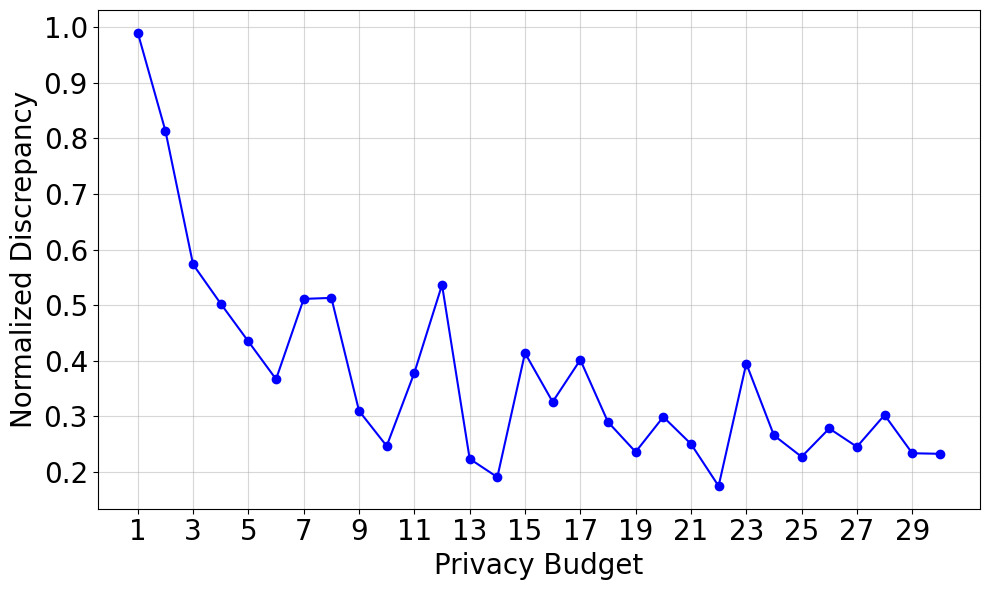} 
        \label{subfig:core500}
    }
    \hfill
    \subfigure[
100-core decomposition]{%
        \includegraphics[width=0.47\textwidth]{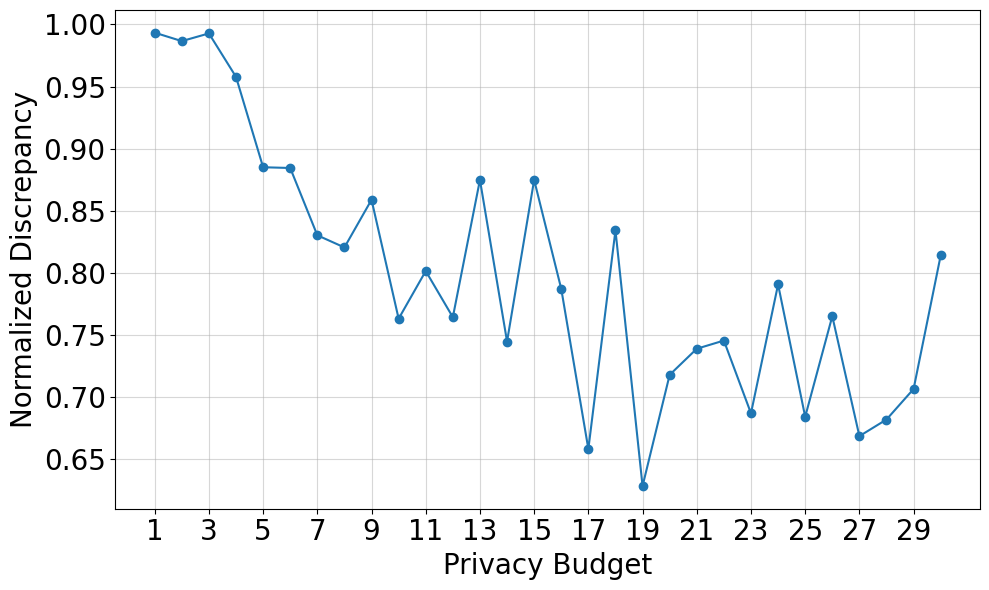} 
        \label{subfig:core100}
    }
    \caption{
The normalized discrepancies of our algorithm for the graph extracted from the Reddit graph}
    \label{fig:overallfigure2}
\end{figure}

Our results for these graphs are shown in Figure~\ref{fig:overallfigure2}. As noted in Section~\ref{subsec:reddit}, clustering these Reddit graphs is more challenging due to the presence of more than two clusters. Even so, for the 500-core decomposition, our algorithm achieves a normalized discrepancy below 0.5 when the privacy budget is at least 4, indicating that it correctly clusters at least half of the vertices.

In the case of the 100-core decomposition, although the discrepancy remains above 0.5 across all privacy budgets, it is still significantly lower than 1— the typical value for random clustering. This suggests that our algorithm performs better than a random baseline when the privacy budget exceeds~4.

\subsection{Results on a typical real graph with more than two clusters: OGBN-Proteins dataset}
We run our algorithm and compare it against randomized response for the Proteins dataset of the Open Graph Benchmark \citep{hu2020opengraphbenchmark}.
In this dataset, nodes represent proteins, and edges indicate biological interactions between them.
The properties about this dataset and its different core decompositions are given by,
\begin{center}
\begin{tabular}{|c|c|c|c|c|}
     \hline Core degree & Number of nodes & Number of edges & $g = \frac{\lambda_2(B)+1}{\lambda_3(B)+1}$ & $T=\frac{2\log n}{\log g}$\\ \hline
     Original & 132,534 & 39,561,252 & 1.001 & 15,939\\
     300      & 64,726  & 29,892,872 & 1.014 & 1,564\\
     500      & 26,136  & 14,887,668 & 1.006 & 3,486\\ \hline
\end{tabular}
\end{center}

\begin{figure}[t]
    \centering
    \subfigure[300-core decomposition]{%
        \includegraphics[width=0.47\textwidth]{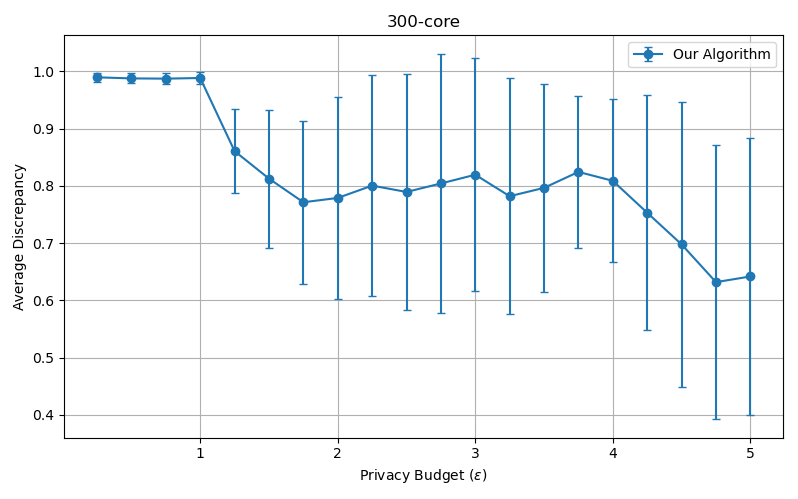} 
        \label{subfig:protein-core300}
    }
    \subfigure[
500-core decomposition]{%
        \includegraphics[width=0.47\textwidth]{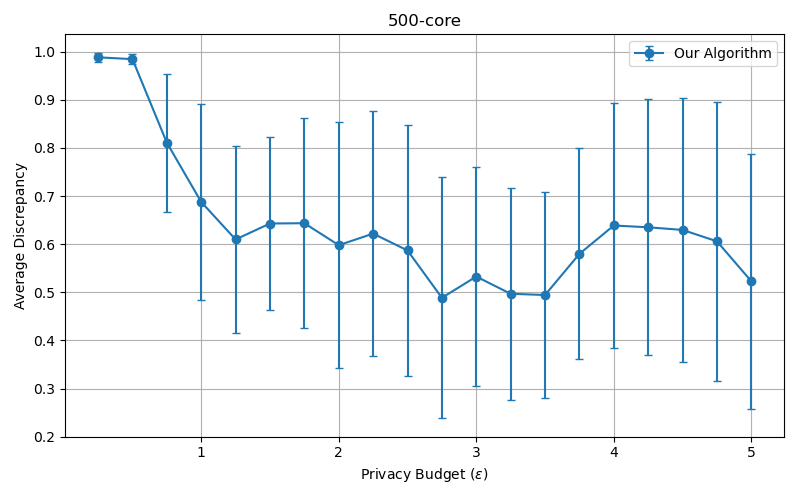} 
        \label{subfig:protein-core500}
    }
    \caption{
The normalized discrepancies of our algorithm for the graphs extracted from the OGBN-Proteins dataset}
    \label{fig:proteinsfull}
\end{figure}

Due to the huge number of vertices in the $300$ and $500$-cores of this graph, we were unable to run the randomized response algorithm even on an A100 GPU with 80 GB of VRAM.
On the other hand, for each of these graphs, due to the absence of a good bicluster structure and very small eigenvalue gap $g$, the number of iterations $T=2\log n/\log g$ incurs too much noise, leading to unstable results. This is why we fix our number of iterations to $50$, as we did for the Reddit graph.
Figure~\ref{fig:proteinsfull} shows the average normalized discrepancy of $20$ runs of our algorithm for the 300 and 500-core decompositions of this network.

We observe similar results as with the Reddit graphs. Since the OGBN-Proteins dataset contains more than two clusters, the eigenvalue gap is close to $1$, which limits our algorithm’s ability to fully recover the clustering structure. Nevertheless, our method successfully identifies substantial portions of the clusters in both networks. When the privacy budget is sufficiently large, approximately half of the nodes can be correctly clustered on average.


\end{document}